\documentclass[12pt, draftclsnofoot, onecolumn]{IEEEtran}
\usepackage{amsmath,amsthm}
\usepackage{nccmath}
\usepackage[utf8]{inputenc} %unicode support
\usepackage{epsfig,scalefnt,multirow}
\usepackage{url}
\usepackage{ifthen}
\usepackage{mathtools}
\usepackage{cite}
\usepackage{graphicx}
\usepackage{amssymb}
\usepackage{tabularx}
\usepackage{amsmath}
\usepackage{epstopdf}
\usepackage{algorithm}

\usepackage{algpseudocode}
\usepackage{algpascal}
\usepackage{cases}
\usepackage{stfloats}
\usepackage{xcolor}
\usepackage{tabularx}% http://ctan.org/pkg/tabularx
\usepackage{soul}
\usepackage{color}
\usepackage{relsize}
\usepackage{booktabs}
\sethlcolor{green}
\usepackage{epsfig,scalefnt,multirow}
\usepackage{url}
\usepackage{ifthen}
\usepackage{mathtools}
\usepackage{cite}
\usepackage{graphicx}
\usepackage[caption=false]{subfig}
\usepackage{graphicx}
\usepackage{amssymb}
\usepackage{booktabs}
\usepackage{amsmath}
\usepackage{epstopdf}
\usepackage{algorithm}
\usepackage{algpseudocode}
\usepackage{algpascal}
\usepackage{cases}
\usepackage{xcolor}

  \def\cC{{\mathcal{C}}}

 \def\cN{{\mathcal{N}}}

\def\bSigma{{\pmb{\Sigma}}} 
\def\bPhi{{\pmb{\Phi}}}

\def\b0{{\pmb{0}}}\def\bLambda{{\pmb{\Lambda}}} 

\def\ba{{\mathbf{a}}}

 \def\bn{{\mathbf{n}}}  
\def\bq{{\mathbf{q}}}  \def\bs{{\mathbf{s}}} 
 \def\bv{{\mathbf{v}}}  \def\bx{{\mathbf{x}}}
\def\by{{\mathbf{y}}}   

\def\bA{{\mathbf{A}}}  \def\bC{{\mathbf{C}}} 
 \def\bF{{\mathbf{F}}} \def\bG{{\mathbf{G}}} \def\bH{{\mathbf{H}}}
\def\bI{{\mathbf{I}}}   
   \def\bP{{\mathbf{P}}}
 \def\bR{{\mathbf{R}}}  
\def\bU{{\mathbf{U}}} \def\bV{{\mathbf{V}}} \def\bW{{\mathbf{W}}}

\DeclarePairedDelimiter\norm{\lVert}{\rVert}

\newtheoremstyle{colon}%
{}
{}
{}%bodyfont
{10pt}%indent
{\itshape}%headfont
{:}%head punctuation
{.5em}%space after head
{\thmname{#1}\thmnumber{ #2}}%

\theoremstyle{colon}

\newtheorem{lemma}{Lemma}
\newtheorem*{lemma*}{Lemma}

\newtheorem{proposition}{Proposition}

\newcommand*{\QEDB}{\null\nobreak\hfill\ensuremath{\square}}%
\ifCLASSINFOpdf
\else
\fi
\begin{document}
	
	\title{Hybrid Beamforming for Intelligent Reflecting Surface Aided Millimeter Wave MIMO Systems}
	
	\author{Sung~Hyuck~Hong, Jaeyong~Park, Sung-Jin~Kim, and~Junil~Choi
		\thanks{Sung Hyuck Hong and Junil Choi are with the School of Electrical Engineering, Korea Advanced Institute of Science and Technology, Daejeon 34141, South Korea (e-mail: \{shong16; junil\}@kaist.ac.kr).}
		\thanks{Jaeyong Park and Sung-Jin Kim are with C\&M Standard Lab, Future Technology Center, LG Electronics Inc. (e-mail: \{jaeyong630.park; sj88.kim\}@lge.com).}}

\maketitle

	\vspace*{-1em}
\begin{abstract}
As communication systems that employ millimeter wave (mmWave) frequency bands must use large antenna arrays to overcome the severe propagation loss of mmWave signals, hybrid beamforming has been considered as an integral component of mmWave communications. Recently, intelligent reflecting surface (IRS) has been proposed as an innovative technology that can significantly improve the performance of mmWave communication systems through the use of low-cost passive reflecting elements. In this paper, we study IRS-aided mmWave multiple-input multiple-output (MIMO) systems with hybrid beamforming architectures. We first exploit the sparse-scattering structure and large dimension of mmWave channels to develop the joint design of IRS reflection matrix and hybrid beamformer for narrowband MIMO systems. Then, we generalize the proposed joint design to broadband MIMO systems with orthogonal frequency division multiplexing (OFDM) modulation by leveraging the angular sparsity of frequency-selective mmWave channels. Simulation results demonstrate that the proposed joint designs can significantly enhance the spectral efficiency of the systems of interest and achieve superior performance over the existing designs.
\end{abstract}


% \fi
\IEEEpeerreviewmaketitle

\section{Introduction}
	\IEEEPARstart{T}{he} explosive growth in wireless data traffic and resultant issue of bandwidth shortage have increased the necessity of resorting to higher frequency bands that are relatively uncongested \cite{PI11}. Since large bandwidth can be utilized at millimeter wave (mmWave) frequency bands to attain possibly up to gigabit-per-second data rates, mmWave communications have been viewed as one of the key technologies that can realize many new applications and services that the fifth generation (5G) wireless networks aim to support \cite{NIU15}. However, the propagation properties of mmWave signals, such as severe path loss and atmospheric attenuation, must be carefully examined in order to fully reap the expected benefits of mmWave communications \cite{ANDREWS14,SWINDLEHURST14}.

Massive multiple-input multiple-output (MIMO), the concept of utilizing a large number of antennas at the transceivers, has been proposed as a promising solution to combat the high path loss of mmWave signals and achieve significant capacity gain through simultaneous transmission of multiple data streams \cite{SWINDLEHURST14,LARSSON14}. Despite its numerous advantages, massive MIMO makes the traditional fully-digital beamforming at baseband prohibitively expensive since such processing requires dedicated radio frequency (RF) chain for each antenna \cite{MOLISCH17}. To address such issue, hybrid beamforming architectures that use the combination of a low-dimensional digital baseband beamformer and a high-dimensional analog beamformer have been proposed to increase the energy efficiency of massive MIMO systems, significantly reducing the number of RF chains at the cost of only slight performance degradation \cite{AHMED18}.

Many works have investigated mmWave MIMO communication systems with hybrid beamforming architectures  \cite{AYACH14,YU16,SOHRABI17,NGUYEN17,CONG18,LIN19,CHEN17_2_a,WANG18_a}. In \cite{AYACH14}, the problem of designing hybrid precoders and combiners that maximize spectral efficiency was recast as a sparsity-constrained matrix reconstruction problem, which was subsequently solved using an algorithm based on orthogonal matching pursuit. Manifold optimization (MO)-based algorithms were proposed in \cite{YU16} to tackle the hybrid precoder design problem in both narrowband and broadband mmWave MIMO systems. The authors in \cite{SOHRABI17} presented a unified heuristic algorithm to design beamformers for frequency-selective mmWave channels under fully-connected and partially-connected hybrid architectures. Hybrid beamformer designs that aim to minimize mean square error (MSE) of data streams were proposed in \cite{NGUYEN17,CONG18,LIN19}, while some works considered the use of finite-resolution phase shifters in implementing analog beamformers \cite{CHEN17_2_a,WANG18_a}.

Recently, intelligent reflecting surface (IRS) has been proposed as a cost-effective and innovative technology to smartly reconfigure the wireless propagation environments in a real-time manner, thereby significantly improving the performance of future communication systems \cite{ZHANG_BJORSON_20}. Labeled as one of the key components of sixth generation (6G) wireless networks, IRS is a metasurface consisting of a large number of passive reflecting elements, each of which induces an adjustable phase and/or amplitude shift to the incident electromagnetic waves \cite{WU19,HUANG20}. Since mmWave signals are highly vulnerable to blockages due to the narrow beamwidth, IRS is, along with massive MIMO and hybrid beamforming, expected to play a pivotal role in enhancing the coverage and spectral/energy efficiency of mmWave communication systems \cite{HUANG20,WU20}.

As 6G wireless networks aim to provide even higher data rates and energy efficiency than 5G networks in which mmWave frequency bands are heavily employed, upcoming communication systems must integrate IRS into mmWave MIMO systems with large antenna arrays and hybrid beamforming architectures to successfully attain such performance targets \cite{YANG19,SAAD20}. However, to reap the full benefits of such integration, it is necessary to overcome the challenge of jointly designing the IRS reflection matrix and hybrid beamformer, both of which are subject to the hardware constraints that must be carefully considered \cite{ZHANG_BJORSON_20}.  Among the limited number of studies on the IRS reflection matrix design for point-to-point MIMO communications \cite{KIM21,PAN20,NING20,ZHANG20,WANG21,YING20}, only \cite{WANG21} and \cite{YING20} considered mmWave MIMO systems with hybrid beamforming architectures. In \cite{WANG21}, a two-stage algorithm based on MO was developed to design the IRS reflection matrix and hybrid beamformer for mmWave systems in which the direct channel from the TX to RX is not present. In \cite{YING20}, the authors constructed the IRS reflection matrix and hybrid beamformer for broadband MIMO systems under frequency-selective channels. However, the hybrid beamformer design proposed in \cite{YING20} focused on minimizing bit error rate (BER) and thus entails an inevitable loss in spectral efficiency. Furthermore, the designs in [26] and [27] construct the hybrid beamformer without considering the effect of the IRS reflection matrix on mmWave channels, although doing so can lead to significant performance improvement and complexity reduction.

Motivated by these facts, we propose in this paper the joint designs of IRS reflection matrix and hybrid beamformer for narrowband and broadband mmWave MIMO systems. The proposed designs aim to maximize the spectral efficiency of IRS-aided mmWave MIMO systems in which the reflected channel from the transmitter (TX) to IRS to receiver (RX) and the direct channel from the TX to RX coexist. To the best of our knowledge, no prior work on IRS-aided mmWave MIMO systems with hybrid beamforming architectures has considered the presence of the direct channel. Furthermore, by exploiting how the IRS reflection matrix influences the structure of mmWave channels when constructing the hybrid beamformer, the proposed joint designs achieve significant performance gains over the existing benchmarks. Our main contributions are summarized as follows:

\begin{itemize}
	\item We develop an IRS reflection matrix design that successfully establishes the favorable communication environments for narrowband mmWave MIMO systems. Then, to demonstrate the effectiveness of the design, we provide an asymptotic analysis of the effect of the proposed IRS reflection matrix on the structure of mmWave channels.
	\item We propose a hybrid beamformer design that takes into account the inherent structure of mmWave channels so as to attain the performance close to that of fully-digital beamforming. By carefully examining how the channels are adjusted according to the proposed IRS reflection matrix design, we construct hybrid precoders and combiners that fully reap considerable spectral efficiency gains offered by IRS.
	\item We generalize the proposed joint design of IRS reflection matrix and hybrid beamformer for narrowband MIMO systems to broadband MIMO systems with orthogonal frequency division multiplexing (OFDM) modulation. Smartly leveraging the sparsity of frequency-selective mmWave channels in the angular domain, the generalized design can significantly enhance the spectral and energy efficiency of IRS-aided MIMO-OFDM systems.  
\end{itemize}

The rest of the paper is organized as follows. The system model and problem formulation are described in Section \ref{sec:SystemModel}. The joint design of IRS reflection matrix and hybrid beamformer for narrowband MIMO systems is proposed in Section \ref{sec:jointdesign}. In Section \ref{sec:extensiontoOFDM}, the proposed joint design is extended to broadband MIMO-OFDM systems. The complexity analysis of the joint designs and simulation results are presented in Section \ref{section:DandCA} and Section \ref{section:simresult}, respectively. Finally, conclusions are drawn in Section \ref{sec:conclusions}.

\textbf{Notation:} Vectors and matrices are represented by lower and upper boldface letters. The $m \times n$ matrix whose elements are all zero is represented by $\b0^{m \times n}$, while ${\mathbb{C}}^{m \times n}$ denotes the set of all $m \times n$ complex matrices. The transpose, conjugate transpose, inverse, determinant, and rank of the matrix $\bA$ are represented by $\bA^{\mathrm{T}}$, $\bA^{\mathrm{H}}$, $\bA^{-1}$, $\text{det}(\bA)$, and $\text{rank}(\bA)$, respectively. The Frobenius norm of $\bA$ is denoted by $\norm{\bA}_{\mathrm{F}}$, while $\text{row}(\bA)$ and $\text{col}(\bA)$ indicate the number of rows and columns of $\bA$, respectively. The element in the $i$-th row and $j$-th column of $\bA$ is denoted by $[\bA]_{i,j}$, whereas $[\bA]_{:,j}$ represents the $j$-th column vector of $\bA$. Similarly,  $[\bA]_{:,1:j} \in \mathbb{C}^{\text{row}(\bA) \times j}$ and  $[\bA]_{1:j,1:j} \in \mathbb{C}^{j \times j}$ each denote the matrix whose columns are given by the first $j$ columns of $\bA$ and that consisting of the first $j$ rows and columns of $\bA$. The $j$-th element of the vector $\bx$ is denoted by $[\bx]_j$, while $\text{diag}(\bx)$ represents the diagonal matrix that contains the elements of $\bx$ on its main diagonal. The $N \times N$ identity matrix and the expectation operator are represented by $\bI_N$ and $\mathbb{E}[\cdot]$, respectively.  The function $\text{max}(a,0)$ defined for a real number $a$ is denoted by $(a)^{+}$, whereas $|{\cdot}|$ and $\norm{\cdot}_2$ each indicate the absolute value of a scalar and the $\ell_2$-norm of a vector. The complex and real normal distributions with mean $m$ and variance $\sigma^2$ are denoted by $\cC \cN(m,\sigma^2)$ and $ \cN(m,\sigma^2)$, respectively.
\begin{figure}[!t]
	\centering
	\includegraphics[width=7.5cm]{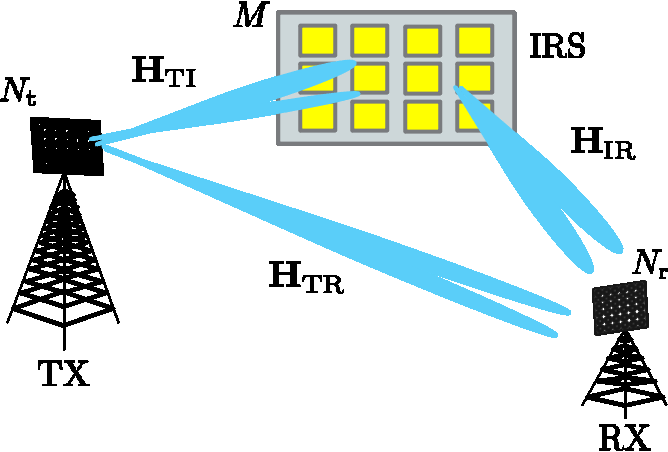}
	%   % where an .eps filename suffix will be assumed under latex,
	%   % and a .pdf suffix will be assumed for pdflatex
	\caption{Illustration of an IRS-aided mmWave MIMO system.}\label{IRS_diagram}
\end{figure}
	\section{System Model and Problem Formulation} \label{sec:SystemModel}
In this paper, we consider an IRS-aided mmWave MIMO system with a hybrid beamforming architecture. For the sake of simplicity, we first describe a narrowband system in this section, and introduce a broadband system later in Section \ref{sec:extensiontoOFDM}. As shown in Fig. \ref{IRS_diagram}, an IRS consisting of $M$ passive reflecting elements assists the communication between the TX with $N_\text{t}$ antennas and the RX with $N_\text{r}$ antennas. Let $\bH_\text{TR} \in {\mathbb{C}}^{N_\text{r} \times N_\text{t}}$ denote the direct channel from the TX to RX. Similarly, let  $\bH_\text{TI} \in {\mathbb{C}}^{M \times N_\text{t}}$  and $\bH_\text{IR} \in {\mathbb{C}}^{N_\text{r} \times M}$ represent the channel from the TX to IRS and that from the IRS to RX, respectively. With each reflecting element of the IRS serving as a single point source that scatters the received signal after applying to it a controllable phase shift, the effect of IRS can be modeled by the diagonal matrix $\bPhi=\text{diag}([e^{j\theta_1},\dots,e^{j\theta_M}]) \in \mathbb{C}^{M \times M}$, where $\theta_m \in [0,2\pi)$ denotes the phase shift of $m$-th IRS element, $m \in \{1,\dots,M\}$ \cite{ZHANG_BJORSON_20,WU19}. The total combined channel from the TX to RX can then be expressed as $\bH_\text{tot}=\bH_\text{TR}+\bH_\text{IR}\bPhi\bH_\text{TI}$.
	
	 The TX sends $N_\text{s}$ data streams to the RX using the digital baseband precoder $\bF_\text{BB} \in {\mathbb{C}}^{N_\text{t}^\text{RF} \times N_\text{s}}$ and analog precoder $\bF_\text{RF} \in {\mathbb{C}}^{N_\text{t} \times N_\text{t}^\text{RF}}$, where the number of RF chains at the TX is denoted by $N_\text{t}^\text{RF}$ and is subject to the constraint $N_\text{s}\leq N_\text{t}^\text{RF} \leq N_\text{t}$. We impose the total power constraint $P_\text{TX}$ on the transmit power, i.e., $\norm{\bF_\text{RF}\bF_\text{BB}}_\mathrm{F}^2 \leq P_\text{TX}$. The RX uses the analog combiner $\bW_{\text{RF}} \in \mathbb{C}^{N_\text{r} \times N^{\text{RF}}_\text{r}}$ and digital baseband combiner $\bW_{\text{BB}} \in \mathbb{C}^{N^{\text{RF}}_\text{r} \times N_\text{s}}$ to process the received signal, where the number of RF chains at the RX $N_\text{r}^\text{RF}$ is subject to the constraint $N_\text{s}\leq N_\text{r}^\text{RF} \leq N_\text{r}$. As the analog combiner $\bW_\text{RF}$ and precoder $\bF_\text{RF}$ are implemented with phase shifters, the constant modulus constraint is imposed on each of their elements, i.e., $|[\bW_\text{RF}]_{m,n}|=1/\sqrt{N_\text{r}}, |[\bF_\text{RF}]_{m,n}|=1/\sqrt{N_\text{t}},\medspace \forall m,n$. The processed received signal is expressed as 
	\begin{align}
	\tilde{\by}&=\bW^{\mathrm{H}}_\text{BB}\bW^{\mathrm{H}}_\text{RF}\bH_\text{tot}\bF_\text{RF} \bF_\text{BB} \bs + \bW^{\mathrm{H}}_\text{BB}\bW^{\mathrm{H}}_\text{RF}\bn, \label{processed_received_signal}
	\end{align}
 where $\bs \in {\mathbb{C}}^{N_\text{s} \times 1}$ is the symbol vector satisfying $\mathbb{E}[\bs\bs^\mathrm{H}]=\bI_{N_\text{s}}$, and $\bn \in {\mathbb{C}}^{N_\text{r} \times 1}$ is an additive white Gaussian noise (AWGN) vector whose entries are independently and identically distributed (i.i.d) with $\cC \cN(0,\sigma^2_{n})$. To accurately evaluate the effectiveness of the proposed IRS reflection matrix and hybrid beamformer design, we assume in this paper that perfect channel state information (CSI) is available at the TX and RX \cite{AYACH14,YU16,SOHRABI17,NING20,ZHANG20,WANG21,YING20}. CSI acquisition at IRS-aided mmWave systems is currently a topic of active research. Recently, several algorithms have been proposed in \cite{WANG_SPL,WANG_TWC,CHEN_TWC} to efficiently estimate mmWave channels in IRS-aided MIMO systems. The achievable spectral efficiency when the transmitted symbols follow a Gaussian distribution is given by
	\begin{align}
	R&=\log_2\text{det}(\bI_{N_\text{s}}+\bR_{\bar{\bn}}^{-1}\bW^\mathrm{H}_\text{BB}\bW^\mathrm{H}_\text{RF}\bH_\text{tot}\bF_\text{RF}\bF_\text{BB}\bF^\mathrm{H}_\text{BB}\bF^\mathrm{H}_\text{RF}\bH_\text{tot}^\mathrm{H} \bW_\text{RF}\bW_\text{BB}),
	\label{spectral_efficiency}
	\end{align}
where $\bR_{\bar{\bn}}=\sigma^2_{n}\bW^\mathrm{H}_\text{BB}\bW^\mathrm{H}_\text{RF}\bW_\text{RF}\bW_\text{BB}$ represents the covariance matrix of the noise $\bar{\bn}=\bW^{\mathrm{H}}_\text{BB}\bW^{\mathrm{H}}_\text{RF}\bn$ after combining.
	
Throughout this paper, we adopt the widely used Saleh-Valenzuela model \cite{AYACH14,YU16,NGUYEN17,CONG18,LIN19,CHEN17_2_a,WANG21,YING20} to represent mmWave channels. Each of the narrowband mmWave channels is expressed as 
	\begin{align}
	\bH_i=\sum_{q=0}^{N_\text{path}^i-1}\alpha_{i,q}\ba_\text{r}(\phi^\text{r}_{i,q},\theta^\text{r}_{i,q})\ba_\text{t}(\phi^\text{t}_{i,q},\theta^\text{t}_{i,q})^{\mathrm{H}}, \label{channel_model}
	\end{align}
	where $i \in \{\text{TR, TI, IR}\}$ is the subscript for the channel matrices, $N_\text{path}^i$ is the number of physical propagation paths in $\bH_i$, and $\alpha_{i,q}$ is the complex gain of the $q$-th path in $\bH_i$. We assume that $\alpha_{i,q}$ are independently distributed with $\cC\cN(0,\gamma_i^2 10^{-0.1PL(d_i)}), \forall q \in \{0,\dots,N_\text{path}^i-1\}$, where $\gamma_i=\sqrt{\text{row}(\bH_i)\text{col}(\bH_i)/N_\text{path}^i}$ is the normalization factor, and $PL(d_i)$ represents the path loss that depends on the distance $d_i$ between the two entities associated with $\bH_i$ \cite{AKDENIZ14}. For example, $d_\text{TR}$ denotes the distance between the TX and RX. Lastly, the normalized receive and transmit array response vectors corresponding to the $q$-th path in $\bH_i$ are respectively denoted by
	$\ba_\text{r}(\phi^\text{r}_{i,q},\theta^\text{r}_{i,q}) \in \mathbb{C}^{\text{row}(\bH_i) \times 1}$ and $\ba_\text{t}(\phi^\text{t}_{i,q},\theta^\text{t}_{i,q}) \in \mathbb{C}^{\text{col}(\bH_i) \times 1}$, where $\phi^\text{r}_{i,q} (\theta^\text{r}_{i,q}$) and  $\phi^\text{t}_{i,q} (\theta^\text{t}_{i,q})$ each stand for the azimuth (elevation) angles of arrival and departure (AoAs and AoDs) of the path. 
	
	We assume in this paper that uniform planar arrays (UPAs) are employed at the TX, RX, and IRS. For the sake of clarity, set $i=\text{TR}$ and define $s \in \{0,\dots,N_\text{path}^\text{TR}-1\}$. The transmit array response vector $\ba_\text{t}(\phi^\text{t}_{\text{TR},s},\theta^\text{t}_{\text{TR},s}) \in \mathbb{C}^{N_\text{t} \times 1}$ corresponding to the $s$-th path in $\bH_\text{TR}$ is then given by
	\begin{align}
	&\ba_\text{t}(\phi^\text{t}_{\text{TR},s},\theta^\text{t}_{\text{TR},s})=\frac{1}{\sqrt{N_\text{t}}}\Big[
	1,\dots,e^{j\frac{2\pi d}{\lambda}(i_h\sin(\phi^\text{t}_{\text{TR},s})\sin(\theta^\text{t}_{\text{TR},s})+i_v\cos(\theta^\text{t}_{\text{TR},s}))},\notag\\
	&\hspace{180pt}\dots,e^{j\frac{2\pi d}{\lambda}((N^h_\text{t}-1)\sin(\phi^\text{t}_{\text{TR},s})\sin(\theta^\text{t}_{\text{TR},s})+(N^v_\text{t}-1)\cos(\theta^\text{t}_{\text{TR},s}))} \Big]^\mathrm{T}, \label{array_example}
	\end{align}
	where $\lambda$ is the signal wavelength, and $d$ is the spacing between the antennas or IRS elements. The horizontal and vertical indices for the transmit antennas are respectively denoted by $0\leq i_h < N^h_\text{t}$ and $0\leq i_v < N^v_\text{t}$, where $N_\text{t}=N^h_\text{t}N^v_\text{t}$. Other receive and transmit array response vectors can be similarly defined.

	In this paper, we aim to maximize the spectral efficiency in (\ref{spectral_efficiency}) by jointly designing the IRS reflection matrix $\bPhi$, hybrid precoder $\bF_\text{RF}\bF_\text{BB}$, and hybrid combiner $\bW_\text{RF}\bW_\text{BB}$. The problem of maximizing the spectral efficiency is formulated as 
	\begin{subequations}
	\begin{align}
	&\text{(P1)}\hspace{42pt}\max_{\bW_\text{BB},\bW_\text{RF},\bPhi,\bF_\text{BB},\bF_{\text{RF}}} R \\
	&\hspace{10pt}\text{subject to } \hspace{13pt}\bPhi=\text{diag}([e^{j\theta_1},\dots,e^{j\theta_M}]) \label{mod_constraint_IRS},\\
	&\hspace{74pt}	\norm{\bF_\text{RF}\bF_\text{BB}}_F^2 \leq P_\text{TX}, \label{TX_power_constraint}\\
	&\hspace{74pt}|[\bW_\text{RF}]_{m,n}|=1/\sqrt{N_\text{r}}, \medspace \forall m,n,\label{mod_constraint_comb}\\ 
	&\hspace{74pt}|[\bF_\text{RF}]_{m,n}|=1/\sqrt{N_\text{t}}, \medspace \forall m,n.\label{mod_constraint_prec} 
	\end{align}
\end{subequations}
	It is challenging to obtain the solution of the optimization problem (P1) since the constant modulus constraints (\ref{mod_constraint_IRS}), (\ref{mod_constraint_comb}), (\ref{mod_constraint_prec}) on $\bPhi$, $\bW_{\text{RF}}$, and $\bF_{\text{RF}}$ are non-convex. Additionally, the objective function $R$, which is coupled with the five matrix variables ${\{\bW_\text{BB},\bW_\text{RF},\bPhi,\bF_\text{BB},\bF_{\text{RF}}\}}$, is neither convex nor concave and thus makes the problem (P1) intractable to solve. To tackle these challenges, we reformulate the problem (P1) by exploiting its structure in the next section.  
	
	\section{Joint Design of IRS Reflection Matrix and Hybrid Beamformer} \label{sec:jointdesign}
	In this section, we propose the joint design of IRS reflection matrix and hybrid beamformer for narrowband mmWave MIMO systems described in Section \ref{sec:SystemModel}. Effectively exploiting the angular sparsity and large dimension of mmWave MIMO channels, the proposed design provides the systems of interest with significant increases in spectral efficiency.
	
	\subsection{Formulation of Effective Channel Design Problem}\label{JointDesign_probForm}
		In this subsection, we transform (P1) into the effective channel design problem, whose purpose is to properly construct the analog combiner $\bW_{\text{RF}}$, IRS reflection matrix $\bPhi$, and analog precoder $\bF_{\text{RF}}$ so that the effective channel $\bH_\text{eff}=\bW^\mathrm{H}_{\text{RF}}\bH_\text{tot}\bF_\text{RF}$ is capable of supporting high spectral efficiency. We first re-express the objective function $R$ of (P1) as
	\begin{align}
	R&=\log_2\text{det}\left(\bI_{N_\text{s}}+\bR_{\bar{\bn}}^{-1}\bW^\mathrm{H}_\text{BB}\bH_\text{eff}\bF_\text{BB}\bF_\text{BB}^\mathrm{H}\bH_\text{eff}^\mathrm{H}\bW_\text{BB} \right),
	\label{reformulation_obj} 
	\end{align}
	where $\bH_\text{eff}=\bW^\mathrm{H}_{\text{RF}}\bH_\text{tot}\bF_\text{RF}$ represents the effective channel after analog precoding and combining are applied. The expression in (\ref{reformulation_obj}) shows that, when $\bH_\text{eff}$ is given, i.e., when $\bPhi, \bW_\text{RF},$ and $\bF_\text{RF}$ are chosen to satisfy the constraints in (\ref{mod_constraint_IRS}), (\ref{mod_constraint_comb}), and (\ref{mod_constraint_prec}), the problem (P1) can be simplified as
	\begin{subequations}
		\begin{align}
		&(\text{P2})\hspace{5pt}  \max_{\bW_\text{BB}, \bF_\text{BB}}\log_2\text{det}(\bI_{N_\text{s}}+\bR_{\bar{\bn}}^{-1}\bW_\text{BB}^\mathrm{H} \bH_\text{eff} \bF_\text{BB} \bF_\text{BB}^\mathrm{H} \bH_\text{eff}^\mathrm{H} \bW_\text{BB}) \\
		&\hspace{22pt}\text{subject to } \hspace{35pt}\norm{\bF_\text{RF}\bF_\text{BB}}_\mathrm{F}^2 \leq P_\text{TX} \label{TX_power_constraint_P2}.
		\end{align}
	\end{subequations}
Note that the problem $(\text{P2})$ cannot be solved in general because the power constraint (\ref{TX_power_constraint_P2}) is coupled with $\bF_\text{RF}$ and the noise $\bW_\text{RF}^\mathrm{H}\bn$ after analog combining is not necessarily white. However, under the condition that $\bW_\text{RF}^\mathrm{H}\bW_\text{RF}=\bI_{N^\text{RF}_\text{r}}$ and $\bF_\text{RF}^\mathrm{H}\bF_\text{RF}=\bI_{N^\text{RF}_\text{t}}$ hold, the optimal solution of the problem (P2) is given by $\{\bW_\text{BB}=\hat{\bW}_\text{BB},\bF_\text{BB}=\hat{\bF}_\text{BB}\}$, where \cite{FMIMO}
\begin{align}
	\hat{\bW}_\text{BB}=[\bU_\text{eff}]_{:,1:N_\text{s}},\hspace{5pt} \hat{\bF}_\text{BB}=[\bV_\text{eff}]_{:,1:N_\text{s}}\bP^{1/2}_\text{eff} \label{tilde_opt_prec/comb}.
\end{align}
	In (\ref{tilde_opt_prec/comb}), $\bU_\text{eff} \in \mathbb{C}^{N^\text{RF}_\text{r} \times N^\text{RF}_\text{r}}$ and $\bV_\text{eff} \in \mathbb{C}^{N^\text{RF}_\text{t} \times N^\text{RF}_\text{t}}$ each denote the unitary matrices whose columns are the left and right singular vectors of $\bH_\text{eff}$, i.e., the singular value decomposition (SVD) of $\bH_\text{eff}$ is expressed as $\bH_\text{eff}=\bU_\text{eff}\bSigma_\text{eff}\bV^\mathrm{H}_\text{eff}$, where $|[\bSigma_\text{eff}]_{m,m}| \geq |[\bSigma_\text{eff}]_{n,n}|, \forall m,n \in \{1,\dots,\text{min}(N^\text{RF}_\text{r},N^\text{RF}_\text{t})\}$ such that  $m < n$. From now on, unless otherwise stated, we will express the SVD of a matrix such that its singular values are in descending order of their absolute values, just as in $\bSigma_\text{eff}$. The waterfilling power allocation matrix $\bP_\text{eff}^{1/2}$ for $\bH_\text{eff}$ is given by  $\bP_\text{eff}^{1/2}=\text{diag}\left(\big[\sqrt{P_1},...,\sqrt{P_{N_\text{s}}}\medspace\big]\right)$, where $P_l=\left(\frac{1}{\eta}-\frac{\sigma^2_{n}}{|[\bSigma_\text{eff}]_{l,l}|^2}\right)^+, \forall l \in \{1,\dots,N_\text{s}\}$, and $\eta$ is chosen such that $\sum_{l=1}^{N_\text{s}}P_l=P_\text{TX}$. The result in \eqref{tilde_opt_prec/comb} indicates that, when the additional constraint is imposed on (P1) so that each of $\bW_{\text{RF}}$ and $\bF_\text{RF}$ has orthonormal columns, the baseband combiner $\hat{\bW}_\text{BB}$ and precoder $\hat{\bF}_\text{BB}$ maximize the spectral efficiency $R$ in (\ref{spectral_efficiency}) for given $\bW_\text{RF}, \bPhi$, and $\bF_\text{RF}$. In fact, $\hat{\bW}_\text{BB}$ and $\hat{\bF}_\text{BB}$ achieve the maximum spectral efficiency\footnote{From now on, we use $R_\text{max}(\bH)$ to denote the maximum spectral efficiency that can be achieved under the transmit power constraint $P_\text{TX}$ and channel $\bH$ of appropriate dimension.} $R_\text{max}(\bH_\text{eff})$ that can be attained under $\bH_\text{eff}$, i.e., 
\begin{align}
R\Bigr|_{\substack{\bW_\text{BB}=\hat{\bW}_\text{BB}\\\bF_\text{BB}=\hat{\bF}_\text{BB}}}
&=\sum_{l=1}^{N_\text{s}}\log_2\left(1+\frac{P_l}{\sigma^2_{n}}|[\bSigma_\text{eff}]_{l,l}|^2\right) =R_\text{max}(\bH_\text{eff})\label{derivation_obj_effective_channel}.
\end{align}
	We therefore aim to maximize $R_\text{max}(\bH_\text{eff})$ in (\ref{derivation_obj_effective_channel}) by properly designing $\bW_\text{RF}, \bPhi$, and $\bF_\text{RF}$ according to the given channels $\bH_\text{TR},\bH_\text{TI},$ and $\bH_\text{IR}$. That is, we focus on solving the effective channel design problem, which is formulated as
	\begin{subequations}
	\begin{align}
	&\text{(P3)}\hspace{57pt} \max_{\bW_\text{RF},\bPhi,\bF_{\text{RF}}} \medspace R_\text{max}(\bH_\text{eff})  \\ 
	&\hspace{10pt}\text{subject to } \hspace{21pt}\bPhi=\text{diag}([e^{j\theta_1},\dots,e^{j\theta_M}]),\label{mod_constraint_IRS_effprob}\\	
	&\hspace{82pt}|[\bW_\text{RF}]_{m,n}|=1/\sqrt{N_\text{r}}, \medspace \forall m,n,\label{mod_constraint_comb_effprob}\\ 
	&\hspace{82pt}|[\bF_\text{RF}]_{m,n}|=1/\sqrt{N_\text{t}}, \medspace \forall m,n,\label{mod_constraint_prec_effprob} \\
	&\hspace{82pt}\bW_\text{RF}^\mathrm{H} \bW_\text{RF}=\bI_{N_\text{r}^\text{RF}},\bF_\text{RF}^\mathrm{H} \bF_\text{RF}=\bI_{N_\text{t}^\text{RF}}.
	\end{align}
\end{subequations}
Note that, given the transmit power constraint $P_\text{TX}$ and noise power $\sigma^2_{n}$, the value of the objective function $R_\text{max}(\bH_\text{eff})$ of the problem (P3) is solely determined by $\left\{ |[\bSigma_\text{eff}]_{l,l}|^2\right\}_{l=1}^{N_\text{s}}$, which are necessarily the $N_\text{s}$ largest eigenvalues of $\bH_\text{eff}\bH_\text{eff}^\mathrm{H}$.           
 
 \subsection{IRS Reflection Matrix Design} \label{sec:IRS_design_singletap}
Although the effective channel design problem (P3) developed in Section \ref{JointDesign_probForm} involves less variables than the original problem (P1) of interest, it is still challenging to find the optimal solution of (P3) since the IRS reflection matrix $\bPhi$ directly influences the total combined channel $\bH_\text{tot}$, which in turn determines the appropriate analog combiner $\bW_\text{RF}$ and precoder $\bF_\text{RF}$. In order to address this difficulty, we propose in this subsection the IRS reflection matrix design that smartly leverages the structure of mmWave MIMO channels to significantly increase the spectral efficiency that $\bH_\text{tot}$ can support. After adjusting $\bPhi$ according to the proposed IRS design, we will construct $\bW_\text{RF}$ and $\bF_\text{RF}$ that successfully translate the spectral efficiency gain in $\bH_\text{tot}=\bH_\text{TR}+\bH_\text{IR}\bPhi\bH_\text{TI}$ to that in $\bH_\text{eff}=\bW_\text{RF}^\mathrm{H}\bH_\text{tot}\bF_\text{RF}$, details of which are explained in Section \ref{JointDesign_analog}.

Let the SVD of $\bH_\text{tot}$ be expressed as  $\bH_\text{tot}=\bU_\text{tot}\bSigma_\text{tot}\bV^\mathrm{H}_\text{tot}$. Then, 
similar to $R_\text{max}(\bH_\text{eff})$ in \eqref{derivation_obj_effective_channel}, the maximum achievable spectral efficiency $R_\text{max}(\bH_\text{tot})$ under $\bH_\text{tot}$ depends only on $\left\{ |[\bSigma_\text{tot}]_{l,l}|^2\right\}_{l=1}^{N_\text{s}}$, or equivalently the $N_s$ largest eigenvalues of $\bH_\text{tot}\bH^\mathrm{H}_\text{tot}$, for given $P_\text{TX}$ and $\sigma^2_{n}$. To mathematically characterize the eigenvalues of $\bH_\text{tot}\bH^\mathrm{H}_\text{tot}$, we first define the receive array response matrix $\bA^{i}_\text{r}$, complex gain matrix $\bG_{i}$, and transmit array response matrix $\bA^{i}_\text{t}$ for the mmWave channel $\bH_i$ in (\ref{channel_model}) as
	\begin{align}
	\bA^{i}_\text{r}&=\begin{bmatrix}
	\ba_\text{r}(\phi^\text{r}_{i,0},\theta^\text{r}_{i,0}) & \cdots & \ba_\text{r}(\phi^\text{r}_{i,N_\text{path}^i-1},\theta^\text{r}_{i,N_\text{path}^i-1})   
	\end{bmatrix}, \notag\\
	&\hspace{33pt}\bG_{i}=\text{diag}\big([\alpha_{i,0},\dots,\alpha_{i,N_\text{path}^i-1}]\big), \notag\\
	\bA^{i}_\text{t}&=\begin{bmatrix}
	\ba_\text{t}(\phi^\text{t}_{i,0},\theta^\text{t}_{i,0}) & \cdots & \ba_\text{t}(\phi^\text{t}_{i,N_\text{path}^i-1},\theta^\text{t}_{i,N_\text{path}^i-1})   
	\end{bmatrix}.
	\end{align}
	The channel matrix $\bH_\text{i}$ can then be decomposed as 
	\begin{align}
	\bH_i=\bA^{i}_\text{r}\bG_{i}(\bA^{i}_\text{t})^{\mathrm{H}},\label{H_i_decomposition}
	\end{align}
	where we assume, without loss of generality, that $|[\bG_i]_{m,m}| \geq |[\bG_i]_{n,n}|, \forall m,n \in \{1,\dots,N_\text{path}^i\}$ such that $m<n$. We now pay particular attention to $\bH_\text{TR}\bH_\text{TI}^\mathrm{H}$, which can be written as 
\begin{align}
\bH_\text{TR}\bH_\text{TI}^\mathrm{H}=\bA_\text{r}^\text{TR}\bG_\text{TR}(\bA_\text{t}^\text{TR})^\mathrm{H}\bA_\text{t}^\text{TI}\bG_\text{TI}^\mathrm{H}(\bA_\text{r}^\text{TI})^\mathrm{H}. \label{H_TR_H_TI}
\end{align}
Since the AoDs of different propagation paths can be considered as continuous random variables that are independent from one another, it follows that the event $E=\big\{ \phi^\text{t}_{\text{TR},s} \neq \phi^\text{t}_{\text{TI},j}, \theta^\text{t}_{\text{TR},s} \neq \theta^\text{t}_{\text{TI},j}, \forall s \in \{ 0,\ldots,N^\text{TR}_\text{path}-1 \}, \forall j \in \{ 0,\ldots,N^\text{TI}_\text{path}-1\} \big\}$ occurs with probability one \cite{STATINF,WANG21}. Then, by the asymptotic orthogonality of UPA array response vectors \cite{CHEN13}, we have
\begin{align}
(\bA_\text{t}^\text{TR})^\mathrm{H}\bA_\text{t}^\text{TI} \rightarrow \b0^{N^\text{TR}_\text{path} \times N^\text{TI}_\text{path}},\medspace \text{as } N_\text{t} \rightarrow \infty, \label{array_matrix_convergence}
\end{align}
which implies that each element of $\bH_\text{TR}\bH_\text{TI}^\mathrm{H}$ converges to $0$ in the limit of large $N_\text{t}$. Using similar arguments, we can show that 
\begin{align}
\bH_\text{tot}\bH^\mathrm{H}_\text{tot}&=\bH_\text{TR}\bH_\text{TR}^\mathrm{H}+\bH_\text{TR}\bH_\text{TI}^\mathrm{H}\bPhi^\mathrm{H}\bH_\text{IR}^\mathrm{H}+\bH_\text{IR}\bPhi\bH_\text{TI}\bH_\text{TR}^\mathrm{H}+\bH_\text{IR}\bPhi\bH_\text{TI}\bH_\text{TI}^\mathrm{H}\bPhi^\mathrm{H}\bH_\text{IR}^\mathrm{H}\notag\\
&\rightarrow\sum_{s=0}^{N_\text{path}^\text{TR}-1}|\alpha_{\text{TR},s}|^2\ba_\text{r}(\phi^\text{r}_{\text{TR},s},\theta^\text{r}_{\text{TR},s})\ba_\text{r}(\phi^\text{r}_{\text{TR},s},\theta^\text{r}_{\text{TR},s})^\mathrm{H}+\sum_{j=0}^{N_\text{path}^\text{TI}-1}|\alpha_{\text{TI},j}|^2\bq_j\bq_j^\mathrm{H}, \label{H_tot_H_tot_H_approx}
\end{align}
as $N_\text{t}\rightarrow \infty$, where  $\bq_j=\bH_\text{IR}\bPhi\ba_\text{r}(\phi^\text{r}_{\text{TI},j},\theta^\text{r}_{\text{TI},j}), \medspace j \in \{0,\dots,N_\text{path}^\text{TI}-1\}$.

We now present the following lemma to establish the connection between the magnitude of $\bq_j$ and IRS reflection matrix $\bPhi$.
\begin{lemma}
For any $j \in \{0,\dots,N_\text{path}^\text{TI}-1\}$, $\norm{\bq_j}^2_2$ satisfies
\begin{align}
 \norm{\bq_j}^2_2 \leq \lambda_0(\bH_\text{IR}^\mathrm{H}\bH_\text{IR}),\label{inequality_lemma1}
 \end{align}
 where the equality holds if $\bPhi\ba_\text{r}(\phi^\text{r}_{\text{TI},j},\theta^\text{r}_{\text{TI},j})$ is the eigenvector corresponding to the maximum eigenvalue $\lambda_0(\bH_\text{IR}^\mathrm{H}\bH_\text{IR})$ of $\bH_\text{IR}^\mathrm{H}\bH_\text{IR}$, i.e.,    $\bH_\text{IR}^\mathrm{H}\bH_\text{IR}\bPhi\ba_\text{r}(\phi^\text{r}_{\text{TI},j},\theta^\text{r}_{\text{TI},j})=\lambda_0(\bH_\text{IR}^\mathrm{H}\bH_\text{IR})\bPhi\ba_\text{r}(\phi^\text{r}_{\text{TI},j},\theta^\text{r}_{\text{TI},j}).$
\end{lemma} 
\hspace{10pt}\textit{Proof:} See Appendix \ref{appendix_A}. \QEDB

With the aid of Lemma 1, we now show in the following proposition that the IRS reflection matrix $\bPhi$ that asymptotically maximizes $\norm{\bq_j}^2_2$ can be obtained in closed-form.
\begin{proposition}
Let $j \in \{0,\dots,N_\text{path}^\text{TI}-1\}$ and define the vector $\bv \in \mathbb{C}^{M \times 1}$ that satisfies $\bPhi=\text{diag}(\bv)$. Then, when $\bv=M\text{diag}(\ba_\text{r}(\phi^\text{r}_{\text{TI},j},\theta^\text{r}_{\text{TI},j})^\mathrm{H})\ba_\text{t}(\phi^\text{t}_{\text{IR},0},\theta^\text{t}_{\text{IR},0})$, it holds that 
\begin{align}
 \norm{\bq_j}^2_2 \rightarrow \lambda_0(\bH_\text{IR}^\mathrm{H}\bH_\text{IR}),\medspace \text{as } N_\text{r}, M \rightarrow \infty. \label{eq_proposition}
\end{align}
\end{proposition}
\hspace{10pt}\textit{Proof:} Set $j \in \{0,\dots,N_\text{path}^\text{TI}-1\}$ and $\bv=M\text{diag}(\ba_\text{r}(\phi^\text{r}_{\text{TI},j},\theta^\text{r}_{\text{TI},j})^\mathrm{H})\ba_\text{t}(\phi^\text{t}_{\text{IR},0},\theta^\text{t}_{\text{IR},0})$. We can then write
\begin{align}
\bPhi\ba_\text{r}(\phi^\text{r}_{\text{TI},j},\theta^\text{r}_{\text{TI},j})&=\text{diag}(\ba_\text{r}(\phi^\text{r}_{\text{TI},j},\theta^\text{r}_{\text{TI},j}))\bv=\ba_\text{t}(\phi^\text{t}_{\text{IR},0},\theta^\text{t}_{\text{IR},0}). \label{phi_a_r_simplified}
\end{align}
By a similar argument used to derive (\ref{array_matrix_convergence}), we have  for each $m \in \{0,\dots,N_\text{path}^\text{IR}-1\}$ that 
	\begin{align}
	\bH_\text{IR}^\mathrm{H}\bH_\text{IR}\ba_\text{t}(\phi^\text{t}_{\text{IR},m},\theta^\text{t}_{\text{IR},m})&=\bA_\text{t}^\text{IR}\bG_\text{IR}^\mathrm{H}(\bA_\text{r}^\text{IR})^\mathrm{H}\bA_\text{r}^\text{IR}\bG_\text{IR}(\bA_\text{t}^\text{IR})^\mathrm{H}\ba_\text{t}(\phi^\text{t}_{\text{IR},m},\theta^\text{t}_{\text{IR},m}) \notag\\
	&\rightarrow\bA_\text{t}^\text{IR}\bG_\text{IR}^\mathrm{H}\bG_\text{IR}[\bI_{N_\text{path}^\text{IR}}]_{:,m+1}=|\alpha_{\text{IR},m}|^2\ba_\text{t}(\phi^\text{t}_{\text{IR},m},\theta^\text{t}_{\text{IR},m}), \label{Lemma2_combined}
	\end{align}
	as $N_\text{r},M \rightarrow \infty$. Since Lemma 1 states that $\norm{\bq_j}^2_2= \lambda_0(\bH_\text{IR}^\mathrm{H}\bH_\text{IR})$ when $\bPhi\ba_\text{r}(\phi^\text{r}_{\text{TI},j},\theta^\text{r}_{\text{TI},j})$ is the eigenvector of $\bH_\text{IR}^\mathrm{H}\bH_\text{IR}$ associated with $\lambda_0(\bH_\text{IR}^\mathrm{H}\bH_\text{IR})$, we can prove Proposition 1 by showing that  $|\alpha_{\text{IR},0}|^2$ asymptotically becomes the maximum eigenvalue of  $\bH_\text{IR}^\mathrm{H}\bH_\text{IR}$ in the limit of large $N_\text{r}$ and $M$. We first note that, by the inequality  $\text{rank}(\bH^\mathrm{H}_\text{IR}\bH_\text{IR})\leq\text{min}(M,N_\text{path}^\text{IR})$, $\bH^\mathrm{H}_\text{IR}\bH_\text{IR}$ can have at most $N_\text{path}^\text{IR}$ nonzero eigenvalues, counting multiplicities, for sufficiently large $M$ such that $M \geq N_\text{path}^\text{IR}$. 
	In fact, (\ref{Lemma2_combined}) implies that the number of nonzero eigenvalues of  $\bH^\mathrm{H}_\text{IR}\bH_\text{IR}$ approaches  $N_\text{path}^\text{IR}$ as $N_\text{r},M \rightarrow \infty$. Since $|\alpha_{\text{IR},0}|^2 \geq |\alpha_{\text{IR},m}|^2$ for each $m$, we have
\begin{align}
\lambda_0(\bH_\text{IR}^\mathrm{H}\bH_\text{IR}) \rightarrow |\alpha_{\text{IR},0}|^2, \text{ as } N_\text{r},M \rightarrow \infty.
\end{align}
This completes the proof of Proposition 1. \QEDB
 	
To examine how setting $\bPhi$ as described in Proposition 1 modifies the value of $\norm{\bq_p}^2_2, p \in \{0,\dots,N_\text{path}^\text{TI}-1\} \setminus \{j\},$ we first express $\bq_p$ as
	\begin{align}
	\bq_p&=\bH_\text{IR}\bPhi\ba_\text{r}(\phi^\text{r}_{\text{TI},p},\theta^\text{r}_{\text{TI},p})=\bA_\text{r}^\text{IR}\bG_\text{IR}(\bA_\text{t}^\text{IR})^\mathrm{H}\text{diag}(\ba_\text{r}(\phi^\text{r}_{\text{TI},p},\theta^\text{r}_{\text{TI},p}))\bv. \label{q_p}
	\end{align}
	By direct computation, we can express the $(m+1)$-th element of $(\bA_\text{t}^\text{IR})^\mathrm{H}\text{diag}(\ba_\text{r}(\phi^\text{r}_{\text{TI},p},\theta^\text{r}_{\text{TI},p}))\bv$ as
	\begin{align}
	&[(\bA_\text{t}^\text{IR})^\mathrm{H}\text{diag}(\ba_\text{r}(\phi^\text{r}_{\text{TI},p},\theta^\text{r}_{\text{TI},p}))\bv]_{m+1}=\ba(f_{m,p},g_{m,p})^\mathrm{H}\ba(f_{0,j},g_{0,j}), 
	\end{align}
	where $m \in \{0,\dots,N_\text{path}^\text{IR}-1\}, f_{m,p}=\sin(\phi^\text{t}_{\text{IR},m})\sin(\theta^\text{t}_{\text{IR},m})-\sin(\phi^\text{r}_{\text{TI},p})\sin(\theta^\text{r}_{\text{TI},p})$, and $g_{m,p}=\cos(\theta^\text{t}_{\text{IR},m})-\cos(\theta^\text{r}_{\text{TI},p})$. The vector $\ba(f,g) \in \mathbb{C}^{M \times 1}$ is expressed as 
	\begin{align}
	&\ba(f,g)=\frac{1}{\sqrt{M}}\Big[
	1,\dots,e^{j\frac{2\pi d}{\lambda}(m_hf+m_vg)},\dots,e^{j\frac{2\pi d}{\lambda}((M^h-1)f+(M^v-1)g)} \Big]^\mathrm{T}. \label{a_vector}
	\end{align}
	Here, $0 \leq m_h < M^h $ and $0 \leq m_v < M^v$ each denote the horizontal and vertical indices for the IRS elements, where $M=M^hM^v$.
	Since the vector $\ba(f,g)$ has the same structure as an $M \times 1$ UPA array response vector, it follows that each element of $(\bA_\text{t}^\text{IR})^\mathrm{H}\text{diag}(\ba_\text{r}(\phi^\text{r}_{\text{TI},p},\theta^\text{r}_{\text{TI},p}))\bv$ tends to zero as $M \rightarrow~\infty$. We can thus conclude from (\ref{q_p}) that  
	\begin{align}
	\norm{\bq_p}^2_2 \rightarrow 0, \text{ as } M \rightarrow \infty, \label{q_p_norm_zero}
	\end{align}
	for any $p \in \{0,\dots,N_\text{path}^\text{TI}-1\} \setminus \{j\}$.

We now explain in the following proposition how the eigenvalues of $\bH_\text{tot}\bH_\text{tot}^\mathrm{H}$ asymptotically behave when $
\bPhi$ is set as described in Proposition 1.

\begin{proposition}
Let $j^\star \in \{0,\dots,N_\text{path}^\text{TI}-1\}$ and $s \in \{0,\dots,N_\text{path}^\text{TR}-1\}$. When $\bPhi=\text{diag}(\bv)$, $\bv=M\text{diag}(\ba_\text{r}(\phi^\text{r}_{\text{TI},j^\star},\theta^\text{r}_{\text{TI},{j^\star}})^\mathrm{H})\ba_\text{t}(\phi^\text{t}_{\text{IR},0},\theta^\text{t}_{\text{IR},0})$, it holds that 
\begin{align}
&\bH_\text{tot}\bH_\text{tot}^\mathrm{H}\ba_\text{r}(\phi^\text{r}_{\text{IR},0},\theta^\text{r}_{\text{IR},0})\rightarrow|\alpha_{\text{TI},j^\star}|^2|\alpha_{\text{IR},0}|^2\ba_\text{r}(\phi^\text{r}_{\text{IR},0},\theta^\text{r}_{\text{IR},0}), \label{Proposition2_1st_eq}\\
&\bH_\text{tot}\bH_\text{tot}^\mathrm{H}\ba_\text{r}(\phi^\text{r}_{\text{TR},s},\theta^\text{r}_{\text{TR},s})\rightarrow|\alpha_{\text{TR},s}|^2\ba_\text{r}(\phi^\text{r}_{\text{TR},s},\theta^\text{r}_{\text{TR},s}), \label{Proposition2_2nd_eq}
\end{align}
as $N_\text{t},N_\text{r},M \rightarrow \infty$.
\end{proposition}
\hspace{10pt}\textit{Proof:} Set $\bv=M\text{diag}(\ba_\text{r}(\phi^\text{r}_{\text{TI},j^\star},\theta^\text{r}_{\text{TI},{j^\star}})^\mathrm{H})\ba_\text{t}(\phi^\text{t}_{\text{IR},0},\theta^\text{t}_{\text{IR},0})$, $j^\star \in \{0,\ldots,N^\text{TI}_\text{path}-1\}$. According to (\ref{H_i_decomposition}), (\ref{phi_a_r_simplified}), and the asymptotic orthogonality of UPA array response vectors, we can express $\bq_{j^\star}$ as 
\begin{align}
\bq_{j^\star}&=\bA_\text{r}^\text{IR}\bG_\text{IR}(\bA_\text{t}^\text{IR})^\mathrm{H}\ba_\text{t}(\phi^\text{t}_{\text{IR},0},\theta^\text{t}_{\text{IR},0})\rightarrow \alpha_{\text{IR},0}\ba_\text{r}(\phi^\text{r}_{\text{IR},0},\theta^\text{r}_{\text{IR},0}), \text{ as } M \rightarrow \infty. \label{q_j_star_convergence}
\end{align} Combining (\ref{q_p_norm_zero}) and \eqref{q_j_star_convergence} with (\ref{H_tot_H_tot_H_approx}), we have 
\begin{align}
&\bH_\text{tot}\bH^\mathrm{H}_\text{tot} \rightarrow\sum_{s=0}^{N_\text{path}^\text{TR}-1}|\alpha_{\text{TR},s}|^2\ba_\text{r}(\phi^\text{r}_{\text{TR},s},\theta^\text{r}_{\text{TR},s})\ba_\text{r}(\phi^\text{r}_{\text{TR},s},\theta^\text{r}_{\text{TR},s})^{\mathrm{H}} \notag\\
&\hspace{120pt}+|\alpha_{\text{TI},j^\star}|^2|\alpha_{\text{IR},0}|^2\ba_\text{r}(\phi^\text{r}_{\text{IR},0},\theta^\text{r}_{\text{IR},0})\ba_\text{r}(\phi^\text{r}_{\text{IR},0},\theta^\text{r}_{\text{IR},0})^\mathrm{H}, \label{H_tot_H_tot^H_withIRS}
\end{align}
as $N_\text{t},M \rightarrow \infty$. Also, it holds for each $s \in \{0,\ldots,N^\text{TR}_\text{path}-1\}$ that
\begin{align}
\ba_\text{r}(\phi^\text{r}_{\text{TR},s},\theta^\text{r}_{\text{TR},s})^\mathrm{H}\ba_\text{r}(\phi^\text{r}_{\text{IR},0},\theta^\text{r}_{\text{IR},0}) \rightarrow 0, \text { as } N_\text{r}\rightarrow \infty, \label{a_r_and_q_orthogonality}
\end{align}   
since the event $T_s=\big\{ \phi^\text{r}_{\text{TR},s} \neq \phi^\text{r}_{\text{IR},0},\theta^\text{r}_{\text{TR},s} \neq \theta^\text{r}_{\text{IR},0}\big\}$ occurs with probability one. It then follows from (\ref{H_tot_H_tot^H_withIRS}) and (\ref{a_r_and_q_orthogonality}) that 
\begin{align}
&\bH_\text{tot}\bH^\mathrm{H}_\text{tot}\ba_\text{r}(\phi^\text{r}_{\text{IR},0},\theta^\text{r}_{\text{IR},0}) \rightarrow |\alpha_{\text{TI},j^\star}|^2|\alpha_{\text{IR},0}|^2\ba_\text{r}(\phi^\text{r}_{\text{IR},0},\theta^\text{r}_{\text{IR},0}),\text{ as } N_\text{t}, N_\text{r}, M \rightarrow \infty,
\end{align}
which finishes the proof of \eqref{Proposition2_1st_eq}. The proof of \eqref{Proposition2_2nd_eq} can be done in a similar manner. \QEDB

	According to Proposition 2, $|\alpha_{\text{TI},j^\star}|^2|\alpha_{\text{IR},0}|^2$  and $\big\{|\alpha_{\text{TR},s}|^2\big\}_{s=0}^{N_\text{path}^{\text{TR}}-1}$ are the asymptotic eigenvalues of $\bH_\text{tot}\bH^\mathrm{H}_\text{tot}$ when $\bv=M\text{diag}(\ba_\text{r}(\phi^\text{r}_{\text{TI},j^\star},\theta^\text{r}_{\text{TI},{j^\star}})^\mathrm{H})\ba_\text{t}(\phi^\text{t}_{\text{IR},0},\theta^\text{t}_{\text{IR},0})$. As $R_\text{max}(\bH_\text{tot})$ monotonically increases with the eigenvalues of $\bH_\text{tot}\bH^\mathrm{H}_\text{tot}$ for given $P_\text{TX}$ and $\sigma_n^2$, we set the IRS reflection matrix as  $\bPhi^\star=\text{diag}(\bv^\star)$, where  
	\begin{align}
	\bv^\star=M\text{diag}(\ba_\text{r}(\phi^\text{r}_{\text{TI},0},\theta^\text{r}_{\text{TI},{0}})^\mathrm{H})\ba_\text{t}(\phi^\text{t}_{\text{IR},0},\theta^\text{t}_{\text{IR},0}), \label{v_star}
	\end{align}
	so that $|\alpha_{\text{TI},0}|^2|\alpha_{\text{IR},0}|^2$ becomes the eigenvalue of $\bH_\text{tot}\bH^\mathrm{H}_\text{tot}$ in the limit of large $N_\text{t}, N_\text{r}$ and $M$.
	
	To investigate how the structure of mmWave MIMO channels is adjusted by the proposed IRS reflection matrix design, we first express the reflected channel $\bH_\text{IR}\bPhi^\star\bH_\text{TI}$ as 
	\begin{align}
	\bH_\text{IR}\bPhi^\star\bH_\text{TI}&=\bA_\text{r}^\text{IR}\bG_\text{IR}(\bA_\text{t}^\text{IR})^\mathrm{H}\bPhi^\star\bA_\text{r}^\text{TI}\bG_\text{TI}(\bA_\text{t}^\text{TI})^\mathrm{H}. \label{adjusted_cascaded_ch}
	\end{align}
Using the expression in \eqref{a_vector}, we can write the element in the $(m+1)$-th row and $(j+1)$-th column of  $\bC=(\bA_\text{t}^\text{IR})^\mathrm{H}\bPhi^\star\bA_\text{r}^\text{TI}$ as 
\begin{align}
	[\bC]_{m+1,j+1}=\begin{cases}
	\ba_\text{t}(\phi^\text{t}_{\text{IR},m},\theta^\text{t}_{\text{IR},m})^\mathrm{H}\ba_\text{t}(\phi^\text{t}_{\text{IR},0},\theta^\text{t}_{\text{IR},0}) \hspace{8pt}\text{ if } j=0, \\
	\ba(f_{m,j},g_{m,j})^\mathrm{H}\ba(f_{0,0},g_{0,0})\hspace{31pt}\text{if } j \neq 0,
	\end{cases} \label{C_matrix}
\end{align}
for each $m \in \{0,\dots,N_\text{path}^\text{IR}-1\}$ and $j \in \{0,\dots,N_\text{path}^\text{TI}-1\}$. It is evident from \eqref{C_matrix} that the asymptotic behavior of $[\bC]_{m+1,j+1}$ can be characterized as 
\begin{align}
[\bC]_{m+1,j+1} \rightarrow \delta_{m0}\delta_{j0}, \text{ as } M \rightarrow \infty,
\end{align}
where $\delta_{xy}$ is the Kronecker delta function that takes the value of 1 if and only if $x=y$. As a result, the adjusted total channel $\bH_\text{tot}^\star=\bH_\text{TR}+\bH_\text{IR}\bPhi^\star\bH_\text{TI}$ satisfies
\begin{align}
&\bH_\text{tot}^\star \rightarrow \sum_{s=0}^{N_\text{path}^\text{TR}-1}\alpha_{\text{TR},s}\ba_\text{r}(\phi^\text{r}_{\text{TR},s},\theta^\text{r}_{\text{TR},s})\ba_\text{t}(\phi^\text{t}_{\text{TR},s},\theta^\text{t}_{\text{TR},s})^{\mathrm{H}}+\alpha_{\text{TI},0}\alpha_{\text{IR},0}\ba_\text{r}(\phi^\text{r}_{\text{IR},0},\theta^\text{r}_{\text{IR},0})\ba_\text{t}(\phi^\text{t}_{\text{TI},0},\theta^\text{t}_{\text{TI},0})^{\mathrm{H}}, \label{total_channel_star_approx} 
\end{align} 
as $M \rightarrow \infty$. It can thus be concluded that, by asymptotically maximizing the eigenvalue associated with the dominant transmit and receive path pair at the IRS, the proposed IRS design successfully establishes a strong communication link between the TX and RX. Also, since $\alpha_{\text{TI},0}$ and $\alpha_{\text{IR},0}$ are independent, we have
\begin{align}
\mathbb{E}\left[|\alpha_{\text{TI},0}|^2|\alpha_{\text{IR},0}|^2\right] \geq M^2c, \label{eigenvalue_ineq}
\end{align}
where $c=N_\text{t}N_\text{r}10^{-0.1(PL(d_\text{TI})+PL(d_\text{IR}))}/N_\text{path}^\text{TI}N_\text{path}^\text{IR}$. The inequality in \eqref{eigenvalue_ineq} shows that, for a fixed value of $c$, the average magnitude of the asymptotic eigenvalue $|\alpha_{\text{TI},0}|^2|\alpha_{\text{IR},0}|^2$ of $\bH^\star_\text{tot}(\bH^\star_\text{tot})^\mathrm{H}$ scales at least quadratically with $M$. This suggests that the proposed design will offer significant increases in spectral efficiency for IRS-aided mmWave MIMO systems, where the IRS passive elements are expected to be deployed in large numbers thanks to their low costs and hardware simplicity \cite{WU19,WU20}.

	 \subsection{Analog Beamformer Design} \label{JointDesign_analog}
	In this subsection, we propose the analog beamformer design that leverages the structure of the favorably adjusted total channel $\bH_\text{tot}^\star=\bH_\text{TR}+\bH_\text{IR}\bPhi^\star\bH_\text{TI}$. Specifically, we construct the analog precoder $\bF_{\text{RF}}$ and combiner $\bW_\text{RF}$ so that the effective channel $\bW_{\text{RF}}^\mathrm{H}\bH_\text{tot}^\star\bF_\text{RF}$ can asymptotically support the maximum spectral efficiency $R_\text{max}(\bH^\star_\text{tot})$ achievable with fully-digital beamforming. 
	
	With the IRS reflection matrix designed according to Section \ref{sec:IRS_design_singletap}, i.e., $\bPhi=\bPhi^\star$, the effective channel design problem (P3) reduces to
\begin{subequations}
	\begin{align}
	&\text{(P4)}\hspace{45pt} \max_{\bW_\text{RF},\bF_{\text{RF}}} \medspace R_\text{max}(\bW_\text{RF}^\mathrm{H}\bH^\star_\text{tot}\bF_{\text{RF}})  \\ 
	&\hspace{10pt}\text{subject to } \hspace{21pt}|[\bW_\text{RF}]_{m,n}|=1/\sqrt{N_\text{r}}, \medspace \forall m,n,\label{constant_mod_constraint_W_P4}\\ 
	&\hspace{82pt}|[\bF_\text{RF}]_{m,n}|=1/\sqrt{N_\text{t}}, \medspace \forall m,n,\label{constant_mod_constraint_F_P4}\\
	&\hspace{82pt}\bW_\text{RF}^\mathrm{H} \bW_\text{RF}=\bI_{N_\text{r}^\text{RF}},\bF_\text{RF}^\mathrm{H} \bF_\text{RF}=\bI_{N_\text{t}^\text{RF}}. \label{semi_unitary_P4}
	\end{align}
\end{subequations}
Removing the constant modulus constraints \eqref{constant_mod_constraint_W_P4}, \eqref{constant_mod_constraint_F_P4} of the problem (P4), we can formulate the relaxed problem  $(\text{P4}^\prime)$ as 
\begin{subequations}
	\begin{align}
	&(\text{P4}^\prime)\hspace{45pt} \max_{\bW_\text{RF},\bF_{\text{RF}}} \medspace R_\text{max}(\bW_\text{RF}^\mathrm{H}\bH^\star_\text{tot}\bF_{\text{RF}})  \\ 
	&\hspace{10pt}\text{subject to } \hspace{21pt}\bW_\text{RF}^\mathrm{H}  \bW_\text{RF}=\bI_{N_\text{r}^\text{RF}},\bF_\text{RF}^\mathrm{H} \bF_\text{RF}=\bI_{N_\text{t}^\text{RF}}. \label{semi_unitary_constraint_P4}
	\end{align}
\end{subequations}
With the SVD of $\bH_\text{tot}^\star$ expressed as $\bH^\star_\text{tot}=\bU^\star_\text{tot}\bSigma^\star_\text{tot}(\bV^\star_\text{tot})^\mathrm{H}$, it is possible to explicitly obtain the optimal solution of $(\text{P4}^\prime)$, as described in the following proposition. 
\begin{proposition}
	Let  $\{\bW_{\text{RF}},\bF_{\text{RF}}\}$ be a feasible solution to the problem $(\text{P4}^\prime)$. Then, it holds that
	\begin{align}
	R_\text{max}(\bW_\text{RF}^\mathrm{H}\bH^\star_\text{tot}\bF_{\text{RF}}) \leq R_\text{max}(\bH_\text{tot}^\star), \label{inequality_prop3}
	\end{align}
	where the equality holds if $\bW_\text{RF}=[\bU_\text{tot}^\star]_{:,1:N_\text{r}^\text{RF}}$ and $\bF_\text{RF}=[\bV_\text{tot}^\star]_{:,1:N_\text{t}^\text{RF}}$. That is, $\{\hat{\bW}_{\text{RF}}^\star=[\bU_\text{tot}^\star]_{:,1:N_\text{r}^\text{RF}},\hat{\bF}_{\text{RF}}^\star=[\bV_\text{tot}^\star]_{:,1:N_\text{t}^\text{RF}}\}$ is optimal for the problem $(\text{P4}^\prime)$.
\end{proposition}
\hspace{10pt}\textit{Proof:} Suppose there exists a feasible solution $\{\bW_{\text{RF}},\bF_{\text{RF}}\}$ of $(\text{P4}^\prime)$ such that \\ $R_\text{max}(\bW_\text{RF}^\mathrm{H}\bH^\star_\text{tot}\bF_{\text{RF}})>R_\text{max}(\bH_\text{tot}^\star)$. Also, let $\{\hat{\bW}^\prime_\text{BB},\hat{\bF}^\prime_\text{BB}\}$ denote the optimal solution of $(\text{P2})$, with $\bH_\text{eff}$ set as $\bH_\text{eff}=\bW_\text{RF}^\mathrm{H}\bH^\star_\text{tot}\bF_{\text{RF}}$. Then, the spectral efficiency equal to  $R_\text{max}(\bW_\text{RF}^\mathrm{H}\bH^\star_\text{tot}\bF_{\text{RF}})$ can be attained under $\bH_\text{tot}^\star$ by using the hybrid combiner $\bW_\text{RF}\hat{\bW}^\prime_\text{BB}$ and precoder $\bF_\text{RF}\hat{\bF}^\prime_\text{BB}$. This contradicts the definition of $R_\text{max}(\bH_\text{tot}^\star)$ and the proof of the inequality in \eqref{inequality_prop3} is complete. 

We now verify the optimality of $\{\hat{\bW}_{\text{RF}}^\star,\hat{\bF}_{\text{RF}}^\star\}$ for the problem $(\text{P4}^\prime)$ by showing that $R_\text{max}(\hat{\bH}_\text{eff}^\star)=R_\text{max}(\bH_\text{tot}^\star)$, where $\hat{\bH}_\text{eff}^\star=(\hat{\bW}_{\text{RF}}^\star)^\mathrm{H}\bH^\star_\text{tot}\hat{\bF}_{\text{RF}}^\star$. Direct computation reveals that $\hat{\bH}_\text{eff}^\star(\hat{\bH}_\text{eff}^\star)^\mathrm{H}$ is the diagonal matrix that satisfies
\begin{align}
[\hat{\bH}_\text{eff}^\star(\hat{\bH}_\text{eff}^\star)^\mathrm{H}]_{n,n}=\begin{cases}
|[\bSigma^\star_\text{tot}]_{n,n}|^2 \hspace{15pt}&\text{if } 1\leq n \leq N_\text{min}^\text{RF} , \\
0 &\text{otherwise,}
\end{cases} 
\end{align}
where $N_\text{min}^\text{RF}=\text{min}(N_\text{r}^\text{RF},N_\text{t}^\text{RF})$. Since $N_\text{s} \leq N_\text{min}^\text{RF}$, the $N_\text{s}$ largest eigenvalues of $\hat{\bH}_\text{eff}^\star(\hat{\bH}_\text{eff}^\star)^\mathrm{H}$ coincide with those of $\bH^\star_\text{tot}(\bH^\star_\text{tot})^\mathrm{H}$, which finishes the proof of Proposition 3. \QEDB

To develop the analog beamformer design that provides an asymptotically optimal solution of (P4), we will now examine the asymptotic property of $\hat{\bW}_{\text{RF}}^\star$ and $\hat{\bF}_{\text{RF}}^\star$. Let us first focus on $\hat{\bW}_{\text{RF}}^\star=[\bU_\text{tot}^\star]_{:,1:N_\text{r}^\text{RF}}$, whose column vectors are necessarily the eigenvectors of $\bH^\star_\text{tot}(\bH^\star_\text{tot})^\mathrm{H}$. The result in \eqref{H_tot_H_tot^H_withIRS} indicates that the number of nonzero eigenvalues of  $\bH^\star_\text{tot}(\bH^\star_\text{tot})^\mathrm{H}$, counting multiplicities, approaches $N_\text{path}^\text{TR}+1$ as $N_\text{t}, N_\text{r}, M \rightarrow \infty$. Since Proposition 2 states that $\{\ba_\text{r}(\phi^\text{r}_{\text{TR},s},\theta^\text{r}_{\text{TR}})\}_{s=0}^{N_\text{path}^\text{TR}-1}$ and $\ba_\text{r}(\phi^\text{r}_{\text{IR},0},\theta^\text{r}_{\text{IR},0})$ asymptotically become the $N_\text{path}^\text{TR}+1$ eigenvectors associated with the nonzero eigenvalues of $\bH^\star_\text{tot}(\bH^\star_\text{tot})^\mathrm{H}$, we can conclude that, as $N_\text{t}, N_\text{r}, M \rightarrow \infty$, each of the eigenvectors that correspond to the nonzero eigenvalues of $\bH^\star_\text{tot}(\bH^\star_\text{tot})^\mathrm{H}$ and are in the columns of $\hat{\bW}_{\text{RF}}^\star$ can be expressed as a scalar multiple of one of the column vectors of $\hat{\bU}^\star_\text{tot}=\begin{bmatrix}
	\bA^\text{TR}_\text{r} & \ba_\text{r}(\phi^\text{r}_{\text{IR},0},\theta^\text{r}_{\text{IR},0})   \\
	\end{bmatrix} \in \mathbb{C}^{N_\text{r} \times (N_\text{path}^\text{TR}+1)}$. By similar argument, we can also show that each eigenvector that is associated with a nonzero eigenvalue of  $(\bH^\star_\text{tot})^\mathrm{H}\bH^\star_\text{tot}$ and is in one of the columns of  $\hat{\bF}_{\text{RF}}^\star=[\bV_\text{tot}^\star]_{:,1:N_\text{t}^\text{RF}}$ approaches a scalar multiple of one of the column vectors of $\hat{\bV}^\star_\text{tot}=\begin{bmatrix}
\bA^\text{TR}_\text{t} & \ba_\text{t}(\phi^\text{t}_{\text{TI},0},\theta^\text{t}_{\text{TI},0})
\end{bmatrix} \in \mathbb{C}^{N_\text{t} \times (N_\text{path}^\text{TR}+1)}$ in the limit of large $N_\text{t}, N_\text{r}$, and $M$.

Motivated by the above formulation, we now construct the analog precoder $\bF_\text{RF}^\star$ and combiner $\bW_\text{RF}^\star$ that are highly effective in IRS-aided mmWave MIMO systems with a large number of antennas and IRS elements. First, we construct the set $\{[\bH_\text{tot}^\star\bA_\text{t}]_{:,c_u}\}_{u=1}^{N_\text{t}^\text{RF}}$ that consists of the $N_\text{t}^\text{RF}$ column vectors of $\bH_\text{tot}^\star\bA_\text{t}$ with the $N_\text{t}^\text{RF}$ largest $\ell_2$-norm, where $c_u \in \{1,\dots,N_\text{path}^\text{TR}+N_\text{path}^\text{TI}\}$. We then set the analog precoder $\bF_\text{RF}^\star$ as
	\begin{align}
	\bF_\text{RF}^\star=\left[[\bA_\text{t}]_{:,c_1},\dots,[\bA_\text{t}]_{:,c_{N_\text{t}^\text{RF}}}\right]. \label{analog_prec_narrow}
	\end{align} 
	Similarly, we construct the analog combiner $\bW_\text{RF}^\star$ as
	\begin{align}
	\bW_\text{RF}^\star=\left[[\bA_\text{r}]_{:,p_1},\dots,[\bA_\text{r}]_{:,p_{N_\text{r}^\text{RF}}}\right], \label{analog_comb_narrow}
	\end{align} 
	where $\{[(\bH_\text{tot}^\star)^\mathrm{H}\bA_\text{r}]_{:,p_v}\}_{v=1}^{N_\text{r}^\text{RF}}$ is the set of $N_\text{r}^\text{RF}$ column vectors of $(\bH_\text{tot}^\star)^\mathrm{H}\bA_\text{r}$ with the $N_\text{r}^\text{RF}$ largest $\ell_2$-norm, and $p_v \in \{1,\dots,N_\text{path}^\text{TR}+N_\text{path}^\text{IR}\}$.
According to Proposition 3 and the asymptotic property of $\hat{\bW}_{\text{RF}}^\star$ and $\hat{\bF}_{\text{RF}}^\star$ discussed in the preceding paragraph, we see that the effective channel $\bH^{\star}_\text{eff}=(\bW^\star_\text{RF})^\mathrm{H}\bH^{\star}_\text{tot}\bF^\star_\text{RF}$ satisfies
\begin{align}
&R_\text{max}(\bH_\text{eff}^\star)\rightarrow R_\text{max}(\hat{\bH}_\text{eff}^\star)=R_\text{max}(\bH_\text{tot}^\star), \label{R_max_H_eff_and_H_tot}
\end{align} 
as $N_\text{t}, N_\text{r}, M \rightarrow \infty$. Therefore, even with limited number of RF chains, the proposed analog beamformer design successfully constructs the effective channel $\bH^{\star}_\text{eff}$ whose maximum supportable spectral efficiency $R_\text{max}(\bH_\text{eff}^\star)$ is very close to $R_\text{max}(\bH_\text{tot}^\star)$ in the systems of interest, thereby harvesting the substantial spectral efficiency gain provided by the IRS reflection matrix design in Section \ref{sec:IRS_design_singletap}.

\subsection{Baseband Beamformer Design} \label{JointDesign_digital}
In the previous two subsections, we described how to design the IRS reflection matrix and analog beamformer for the channel matrices $\bH_\text{TR},\bH_\text{TI}$ and $\bH_\text{IR}$. In this subsection, we propose the design of the baseband combiner and precoder that are nearly optimal for the effective channel $\bH^{\star}_\text{eff}=(\bW^\star_\text{RF})^\mathrm{H}\bH^{\star}_\text{tot}\bF^\star_\text{RF}$. As explained in Section \ref{JointDesign_probForm}, when $(\bW^\star_\text{RF})^\mathrm{H}\bW^\star_\text{RF}=\bI_{N^\text{RF}_\text{r}}$ and $(\bF^\star_\text{RF})^\mathrm{H}\bF^\star_\text{RF}=\bI_{N^\text{RF}_\text{t}}$ hold, the baseband combiner $\hat{\bW}^\star_\text{BB}$ and precoder $\hat{\bF}^\star_\text{BB}$ that maximize the spectral efficiency in \eqref{reformulation_obj} for given $\bH_\text{eff}=\bH_\text{eff}^\star$  are 
\begin{align}
\hspace{20pt}\hat{\bW}^\star_\text{BB}=[\bU^\star_\text{eff}]_{:,1:N_\text{s}},\hspace{5pt} \hat{\bF}^\star_\text{BB}=[\bV^\star_\text{eff}]_{:,1:N_\text{s}}(\bP^\star_\text{eff})^{1/2},\label{tilde_F_BB_and_W_BB^star}
\end{align}
where the SVD of $\bH_\text{eff}^\star$ is expressed as $\bH_\text{eff}^\star=\bU_\text{eff}^\star\bSigma_\text{eff}^\star(\bV_\text{eff}^\star)^\mathrm{H}$.
The power allocation matrix $(\bP^\star_\text{eff})^{1/2}$ is given by $(\bP^\star_\text{eff})^{1/2}=\text{diag}\left(\big[\sqrt{P^\star_1},...,\sqrt{P^\star_{N_\text{s}}}\big]\right)$, where $P^\star_l=\left(\frac{1}{\eta^\star}-\frac{\sigma^2_{n}}{|[\bSigma^\star_\text{eff}]_{l,l}|^2}\right)^+, \forall l \in \{1,\dots,N_\text{s}\}$, and $\eta^\star$ is chosen such that $\sum_{l=1}^{N_\text{s}}P^\star_l=P_\text{TX}$. As both $\bW^\star_\text{RF}$ and $\bF^\star_\text{RF}$ have UPA array response vectors in their respective columns, $(\bW^\star_\text{RF})^\mathrm{H}\bW^\star_\text{RF}$ and $(\bF^\star_\text{RF})^\mathrm{H}\bF^\star_\text{RF}$ each approach $\bI_{N^\text{RF}_\text{r}}$ and $\bI_{N^\text{RF}_\text{t}}$ as $N_\text{t},N_\text{r} \rightarrow \infty$. Therefore, using $\hat{\bW}^\star_\text{BB}$ and $\hat{\bF}^\star_\text{BB}$ as the baseband combiner and precoder is nearly optimal in mmWave systems, where large antenna arrays are employed at the TX and RX. Note that, while $\bW^\star_\text{RF}\hat{\bW}^\star_\text{BB}$ can be directly used as the hybrid combiner, there is no guarantee that the hybrid precoder $\bF^\star_\text{RF}\hat{\bF}^\star_\text{BB}$ satisfies the transmit power constraint \eqref{TX_power_constraint} of the problem (P1), i.e., $\norm{\bF^\star_\text{RF}\hat{\bF}^\star_\text{BB}}_\mathrm{F}^2$ might be greater than $P_\text{TX}$. With this difference in mind, we adopt $\bW_\text{BB}^\star=\hat{\bW}^\star_\text{BB}$ and $\bF_\text{BB}^\star=\gamma^\star\hat{\bF}^\star_\text{BB}$ as the baseband combiner and precoder, where 
 \begin{align}
\gamma^\star=\frac{\sqrt{P_\text{TX}}}{\|\bF^\star_\text{RF}\hat{\bF}^\star_\text{BB}\|_\mathrm{F}} \label{gamma_star}
 \end{align}
 is the normalization factor that ensures the transmit power constraint is met with equality, i.e., $\|\bF^\star_\text{RF}\bF^\star_\text{BB}\|^2_\mathrm{F}=P_\text{TX}$. Since the asymptotic orthogonality of UPA array response vectors guarantees that $\bF^\star_\text{BB}$ converges to $\hat{\bF}^\star_\text{BB}$ as $N_\text{t}\rightarrow \infty$, we can conclude from \eqref{R_max_H_eff_and_H_tot} that the proposed hybrid precoder $\bF^\star_\text{RF}\bF^\star_\text{BB}$ and combiner $\bW^\star_\text{RF}\bW^\star_\text{BB}$ achieve spectral efficiency that asymptotically approaches $R_\text{max}(\bH_\text{eff}^\star)$, or equivalently $R_\text{max}(\bH_\text{tot}^\star)$, as $N_\text{t}, N_\text{r}, M \rightarrow \infty$. Therefore, in typical IRS-aided mmWave MIMO systems with a large number of antennas and IRS elements, the proposed hybrid beamformer design can perform close to the optimal fully-digital beamformer design, providing the systems with considerable benefits in terms of cost and energy efficiency.

	\section{Extension of Joint Design to MIMO-OFDM Systems} \label{sec:extensiontoOFDM}
As one of the key features of mmWave communications is the usage of large bandwidth, it is important to investigate the design of IRS reflection matrix and hybrid beamformer for frequency-selective mmWave channels. In this section, we extend the proposed joint design of IRS reflection matrix and hybrid beamformer in Section \ref{sec:jointdesign} to broadband MIMO-OFDM systems.

\subsection{System Model and Problem Formulation}
Consider an IRS-aided MIMO-OFDM system where the TX performs digital precoding to the symbol vector $\bs[k] \in \mathbb{C}^{N_\text{s} \times 1}$ in the frequency domain by using the baseband precoder $\bF_\text{BB}[k] \in \mathbb{C}^{N_\text{t}^\text{RF} \times N_\text{s}}$ at each subcarrier $k \in \{0,\dots,K-1\}$. The precoded symbol vector is then transformed into the time domain through the $K$-point inverse fast Fourier transform (FFT) followed by cyclic prefix (CP) addition at each of the $N_\text{t}^\text{RF}$ RF chains. Subsequently, the TX applies the analog precoder $\bF_\text{RF}$ to the transformed vector to produce the final transmitted signal. We assume that $\bs[k]$ satisfies $\mathbb E[\bs[k] \bs[k]^\mathrm{H}]=\bI_{N_\text{s}}$, and that the TX is subject to the per-subcarrier power constraint, i.e., $\norm{\bF_\text{RF}\bF_\text{BB}[k]}_\mathrm{F}^2\leq P_\text{TX}[k]$. The received signal at the subcarrier $k$ can be expressed as
\begin{align}
\by[k]&=(\bH_\text{TR}[k]+\bH_\text{IR}[k]\bPhi\bH_\text{TI}[k])\bF_\text{RF}\bF_\text{BB}[k]\bs[k]+\bn[k] =\bH_\text{tot}[k]\bF_\text{RF}\bF_\text{BB}[k]\bs[k]+\bn[k]
\end{align}
where $\bH_\text{TR}[k] \in \mathbb{C}^{N_\text{r} \times N_\text{t}},\bH_\text{TI}[k] \in \mathbb{C}^{M \times N_\text{t}}$, and $\bH_\text{IR}[k] \in \mathbb{C}^{N_\text{r} \times M}$ denote the frequency-domain channel matrices at the subcarrier $k$ from the TX to RX, from the TX to IRS, and from the IRS to RX, respectively. The total combined channel matrix at the subcarrier $k$ from the TX to RX is denoted by $\bH_\text{tot}[k]=\bH_\text{TR}[k]+\bH_\text{IR}[k]\bPhi\bH_\text{TI}[k]$, and the AWGN vector $\bn[k]$ has entries that are i.i.d with $\cC \cN(0,\sigma^2_{n})$. The RX first applies the analog combiner $\bW_\text{RF}$ to the received signal in the time domain, and then transforms it into the frequency domain by removing the CP and performing the $K$-point FFT at each of its $N_\text{r}^\text{RF}$ RF chains. Finally, the RX uses the baseband combiner $\bW_\text{BB}[k] \in \mathbb{C}^{N_\text{r}^\text{RF} \times N_\text{s}}$ to obtain the processed received signal $\tilde{\by}[k]=\bW_{\text{BB}}[k]^\mathrm{H}\bW_\text{RF}^\mathrm{H}\by[k]$ at each subcarrier $k$. For each $i \in \{\text{TR, TI, IR}\}$, the channel matrix $\bH_i[k]$ at the subcarrier $k$ is expressed as \cite{YU16} 
\begin{align}
\bH_i[k]=\sum_{q=0}^{N_\text{path}^i-1}\alpha_{i,q}\ba_\text{r}(\phi^\text{r}_{i,q},\theta^\text{r}_{i,q})\ba_\text{t}(\phi^\text{t}_{i,q},\theta^\text{t}_{i,q})^{\mathrm{H}}e^{-j 2 \pi qk/K}. \label{mmwave_ch_matrix}
\end{align} 
The achievable spectral efficiency over the subcarrier $k$ is given by
\begin{align}
&R[k]=\log_2\text{det}(\bI_{N_\text{s}}+\bR_{\bar{\bn}[k]}^{-1}\bW_\text{BB}[k]^\mathrm{H}\bW^\mathrm{H}_\text{RF}\bH_\text{tot}[k]\bF_\text{RF}\bF_\text{BB}[k]\bF_\text{BB}[k]^\mathrm{H}\bF^\mathrm{H}_\text{RF}\bH_\text{tot}[k]^\mathrm{H} \bW_\text{RF}\bW_\text{BB}[k]),
\label{spectral_efficiency_k_th_sc}
\end{align}
where $\bR_{\bar{\bn}[k]}=\sigma^2_{n}\bW_\text{BB}[k]^\mathrm{H}\bW^\mathrm{H}_\text{RF}\bW_\text{RF}\bW_\text{BB}[k]$ is the covariance matrix of  $\bar{\bn}[k]=\bW_\text{BB}[k]^\mathrm{H}\bW^\mathrm{H}_\text{RF}\bn[k]$.

The problem of designing the IRS reflection matrix and hybrid beamformer that maximize the spectral efficiency of MIMO-OFDM systems can be formulated as 
\begin{subequations}
		\begin{align}
		&\text{(P5)}\hspace{13pt}\max_{\substack{\hspace{12pt}\{\bW_\text{BB}[k],\bF_\text{BB}[k]\}_{k=0}^{K-1},\\\bW_\text{RF},\bPhi,\bF_{\text{RF}}}}  \hspace{2pt}\sum_{k=0}^{K-1} R[k] \label{obj_func_P5} \\
		&\hspace{4pt}\text{subject to } \hspace{7pt}\bPhi=\text{diag}([e^{j\theta_1},\dots,e^{j\theta_M}]),\\
		&\hspace{54pt}\norm{\bF_\text{RF}\bF_\text{BB}[k]}_\mathrm{F}^2\leq P_\text{TX}[k], \label{power_constraint_OFDM}\\
		&\hspace{54pt}|[\bW_\text{RF}]_{m,n}|=1/\sqrt{N_\text{r}}, \medspace \forall m,n,\\ 
		&\hspace{54pt}|[\bF_\text{RF}]_{m,n}|=1/\sqrt{N_\text{t}}, \medspace \forall m,n. 
		\end{align}
	\end{subequations}
Following the similar steps used to derive the problem (P3) in Section \ref{JointDesign_probForm}, we can formulate the effective channel design problem for MIMO-OFDM systems as 
\begin{subequations}
	\begin{align}
	&\text{(P6)}\hspace{35pt} \max_{\bW_\text{RF},\bPhi,\bF_{\text{RF}}} \medspace \sum_{k=0}^{K-1} R_\text{max}(\bH_\text{eff}[k])  \\ 
	&\hspace{4pt}\text{subject to } \hspace{11pt}\bPhi=\text{diag}([e^{j\theta_1},\dots,e^{j\theta_M}]),\label{mod_constraint_IRS_effprob_OFDM}\\
	&\hspace{58pt}|[\bW_\text{RF}]_{m,n}|=1/\sqrt{N_\text{r}}, \medspace \forall m,n,\label{mod_constraint_comb_effprob_OFDM}\\ 
	&\hspace{58pt}|[\bF_\text{RF}]_{m,n}|=1/\sqrt{N_\text{t}}, \medspace \forall m,n,\label{mod_constraint_prec_effprob_OFDM} \\
	&\hspace{58pt}\bW_\text{RF}^\mathrm{H} \bW_\text{RF}=\bI_{N_\text{r}^\text{RF}},\bF_\text{RF}^\mathrm{H} \bF_\text{RF}=\bI_{N_\text{t}^\text{RF}},
	\end{align}
\end{subequations}
where $R_\text{max}(\bH_\text{eff}[k])$ denotes the maximum achievable spectral efficiency under the effective channel $\bH_\text{eff}[k]=\bW^\mathrm{H}_\text{RF}\bH_\text{tot}[k]\bF_\text{RF}$ at the subcarrier $k$. Note that, due to the lack of baseband processing capabilities at the IRS \cite{YING20,ZHANG20}, the design of $\bPhi$ must take the channel matrices of all the subcarriers into account. Likewise, $\bF_\text{RF}$ and $\bW_\text{RF}$ are set to be identical for all the subcarriers since both of them are applied after the inverse FFT operation. These two characteristics of MIMO-OFDM systems make it impossible to directly apply the joint design proposed in Section \ref{sec:jointdesign} to frequency-selective mmWave channels.
%teleport
\subsection{IRS Reflection Matrix Design for MIMO-OFDM Systems} \label{IRS_design_OFDM}
To tackle the effective channel design problem (P6) for MIMO-OFDM systems, we adopt the strategy similar to the one described in Section \ref{sec:jointdesign}, where we first designed the IRS reflection matrix $\bPhi$, and then constructed the analog precoder $\bF_\text{RF}$ and combiner $\bW_\text{RF}$ accordingly. We first define the complex gain matrix $\bG_{i}[k]$ for the mmWave channel $\bH_i[k]$ in \eqref{mmwave_ch_matrix} as
\begin{align}
\bG_{i}[k]=\text{diag}\left(\big[ \alpha_{i,0},\dots, \alpha_{i,N_\text{path}^i-1}\cdot e^{-j2 \pi (N_\text{path}^i-1)k/K}\big]\right),
\end{align}
where, without loss of generality, we assume that $|[\bG_i[k]]_{m,m}| \geq |[\bG_i[k]]_{n,n}|, \forall m,n \in \{1,\dots,N_\text{path}^i\}$ such that $m<n$. We can then express the decomposition of $\bH_i[k]$ as
$\bH_i[k]=\bA^{i}_\text{r}\bG_{i}[k](\bA^{i}_\text{t})^{\mathrm{H}}$,
which has the identical structure as that of $\bH_i$ in \eqref{H_i_decomposition}. Therefore, by following the similar steps used to derive \eqref{H_tot_H_tot^H_withIRS}, we can show that $\bH^\star_\text{tot}[k]=\bH_\text{TR}[k]+\bH_\text{IR}[k]\bPhi^\star\bH_\text{TI}[k]$ satisfies
\begin{align}
&\bH^\star_\text{tot}[k](\bH^\star_\text{tot}[k])^\mathrm{H}\rightarrow \sum_{s=0}^{N_\text{path}^\text{TR}-1}|\alpha_{\text{TR},s}|^2\ba_\text{r}(\phi^\text{r}_{\text{TR},s},\theta^\text{r}_{\text{TR},s})\ba_\text{r}(\phi^\text{r}_{\text{TR},s},\theta^\text{r}_{\text{TR},s})^{\mathrm{H}} \notag\\
&\hspace{150pt}+|\alpha_{\text{TI},0}|^2|\alpha_{\text{IR},0}|^2\ba_\text{r}(\phi^\text{r}_{\text{IR},0},\theta^\text{r}_{\text{IR},0})\ba_\text{r}(\phi^\text{r}_{\text{IR},0},\theta^\text{r}_{\text{IR},0})^{\mathrm{H}}, \label{H_tot_H_tot_star_k}
\end{align}
as $N_\text{t}, M \rightarrow \infty$, where $\bPhi^\star=\text{diag}(\bv^\star)$ is the IRS reflection matrix proposed in Section \ref{sec:IRS_design_singletap}, and $\bv^\star$ is given in \eqref{v_star}. Since \eqref {H_tot_H_tot^H_withIRS} and \eqref{H_tot_H_tot_star_k} indicate that $\bH^\star_\text{tot}[k](\bH^\star_\text{tot}[k])^\mathrm{H}$ and $\bH^\star_\text{tot}(\bH^\star_\text{tot})^\mathrm{H}$ converge to the same matrix for each $k \in \{0,\dots,K-1\}$, the IRS reflection matrix $\bPhi^\star$ has the same asymptotic effect on the eigenvalues of $\bH_\text{tot}^\star[k](\bH_\text{tot}^\star[k])^\mathrm{H}$ as it does on those of $\bH^\star_\text{tot}(\bH^\star_\text{tot})^\mathrm{H}$. As the inequality in \eqref{eigenvalue_ineq} guarantees that the asymptotic eigenvalue $|\alpha_{\text{TI},0}|^2|\alpha_{\text{IR},0}|^2$ of $\bH^\star_\text{tot}(\bH^\star_\text{tot})^\mathrm{H}$ and $\bH_\text{tot}^\star[k](\bH_\text{tot}^\star[k])^\mathrm{H}$ increases at least quadratically with the number of IRS elements $M$, it can be concluded that the proposed IRS reflection matrix design in Section \ref{sec:IRS_design_singletap} will provide both narrowband and broadband MIMO systems with considerable gains in spectral efficiency.

 \subsection{Analog Beamformer Design for MIMO-OFDM Systems} \label{Joint_Design_analog_OFDM}
In the previous subsection, we showed that every frequency-domain channel matrix $\bH_i[k]$ at the subcarrier $k$ shares the same receive array response matrix $\bA_\text{r}^i$ and transmit array response matrix $\bA_\text{t}^i$. In this subsection, we exploit this particular property of frequency-selective mmWave channels to extend the proposed analog beamformer design in Section \ref{JointDesign_analog} to MIMO-OFDM systems. With $\bPhi=\bPhi^\star$, the effective channel design problem (P6) is simplified as 
\begin{subequations}
	\begin{align}
	&\text{(P7)}\hspace{33pt} \max_{\bW_\text{RF},\bF_{\text{RF}}} \medspace \sum_{k=0}^{K-1}  R_\text{max}(\bW_\text{RF}^\mathrm{H}\bH^\star_\text{tot}[k]\bF_{\text{RF}}) \label{obj_function_P7}  \\ 
	&\hspace{10pt}\text{subject to } \hspace{10pt}|[\bW_\text{RF}]_{m,n}|=1/\sqrt{N_\text{r}}, \medspace \forall m,n,\\ 
	&\hspace{71pt}|[\bF_\text{RF}]_{m,n}|=1/\sqrt{N_\text{t}}, \medspace \forall m,n,\\
	&\hspace{71pt}\bW_\text{RF}^\mathrm{H} \bW_\text{RF}=\bI_{N_\text{r}^\text{RF}},\bF_\text{RF}^\mathrm{H} \bF_\text{RF}=\bI_{N_\text{t}^\text{RF}}. 
	\end{align}
\end{subequations}
Now consider $\bF^{\star}_{\text{RF},k^*} \in \mathbb{C}^{N_\text{t} \times N_\text{t}^\text{RF}}$ and $\bW^{\star}_{\text{RF},k^*} \in \mathbb{C}^{N_\text{r} \times N_\text{r}^\text{RF}}$, which denote the analog precoder and combiner obtained by applying the proposed analog beamformer design in Section \ref{JointDesign_analog} to $\bH^\star_\text{tot}[k^*]$ for some $k^* \in \{0,\dots,K-1\}$. Then, it holds for each $k \in \{0,\dots,K-1\}$ that $R_\text{max}((\bW^{\star}_{\text{RF},k^*})^\mathrm{H}\bH^\star_\text{tot}[k]\bF^{\star}_{\text{RF},k^*})$ asymptotically approaches the maximum achievable spectral efficiency  $R_\text{max}(\bH^\star_\text{tot}[k])$ under $\bH^\star_\text{tot}[k]$. To see this, note that, as discussed in Section \ref{IRS_design_OFDM},
	$\bH^\star_\text{tot}[k](\bH^\star_\text{tot}[k])^\mathrm{H}$ and $\bH^\star_\text{tot}(\bH^\star_\text{tot})^\mathrm{H}$ converge to the same matrix as $N_\text{t},M \rightarrow \infty$. Likewise, it can be shown that $(\bH^\star_\text{tot}[k])^\mathrm{H}\bH^\star_\text{tot}[k]$ and $(\bH^\star_\text{tot})^\mathrm{H}\bH^\star_\text{tot}$ converge to the same $N_\text{t} \times N_\text{t}$ matrix as $N_\text{r}, M\rightarrow \infty$. Furthermore, since Proposition 3 implies that $R_\text{max}(\bW_\text{RF}^\mathrm{H}\bH_\text{tot}^\star[k]\bF_\text{RF}) \leq R_\text{max}(\bH^\star_\text{tot}[k])$ for each feasible solution $\{\bW_\text{RF},\bF_\text{RF}\}$ of the problem (P7), we have that $\{\bW_\text{RF}=\bW^{\star}_{\text{RF},k^*}, \bF_\text{RF}=\bF^{\star}_{\text{RF},k^*}\}$ approaches the optimal solution of (P7) as $N_\text{t}, N_\text{r}, M \rightarrow \infty$.

With these formulations in mind, we now explain the analog beamformer design for frequency-selective mmWave channels. First, we use the analog beamformer design proposed in Section \ref{JointDesign_analog} to create the block matrices $\bF^{\text{block}}_\text{RF} \in \mathbb{C}^{N_\text{t} \times KN_\text{t}^\text{RF}}$ and $\bW^{\text{block}}_\text{RF} \in \mathbb{C}^{N_\text{r} \times KN_\text{r}^\text{RF}}$, where
\begin{align}
\bF^{\text{block}}_\text{RF}&=\begin{bmatrix}
\bF^\star_{\text{RF},0} & \bF^\star_{\text{RF},1}  & \dots & \bF^\star_{\text{RF},K-1} 
\end{bmatrix}, 
\bW^{\text{block}}_\text{RF}&=\begin{bmatrix}
\bW^\star_{\text{RF},0} & \bW^\star_{\text{RF},1}  & \dots & \bW^\star_{\text{RF},K-1}
\end{bmatrix}.
\end{align}
Then, the proposed analog precoder and combiner are given by
	\begin{align}
	\bF_\text{RF}^\star=\left[[\bA_\text{t}]_{:,e_1},\dots,[\bA_\text{t}]_{:,e_{N_\text{t}^\text{RF}}}\right], \label{analog_prec_wideband}\\
	\bW_\text{RF}^\star=\left[[\bA_\text{r}]_{:,f_1},\dots,[\bA_\text{r}]_{:,f_{N_\text{r}^\text{RF}}}\right],\label{analog_comb_wideband}
	\end{align}
	where 
	$\{[\bA_\text{t}]_{:,e_u}\}_{u=1}^{N_\text{t}^\text{RF}}$ is the set of $N_\text{t}^\text{RF}$ vectors that appear most frequently in the columns of $\bF^{\text{block}}_\text{RF}$, $e_u \in \{1,\dots,N_\text{path}^\text{TR}+N_\text{path}^\text{TI}\}$, $\{[\bA_\text{r}]_{:,f_v}\}_{v=1}^{N_\text{r}^\text{RF}}$ is the set of $N_\text{r}^\text{RF}$ vectors that appear most frequently in the columns of $\bW^{\text{block}}_\text{RF}$, and $f_v \in \{1,\dots,N_\text{path}^\text{TR}+N_\text{path}^\text{IR}\}$. Analogous to \eqref{R_max_H_eff_and_H_tot}, we have
	\begin{align}
	\sum_{k=0}^{K-1}R_\text{max}(\bH_\text{eff}^\star[k]) \rightarrow  \sum_{k=0}^{K-1}R_\text{max}(\bH_\text{tot}^\star[k]), \label{eff_channel_spec_convg_to_channel_spec}
	\end{align}
	as $N_\text{t}, N_\text{r}, M \rightarrow \infty$, where $\bH_\text{eff}^\star[k]=(\bW^\star_\text{RF})^\mathrm{H}\bH^\star_\text{tot}[k]\bF^\star_{\text{RF}}$. The result in \eqref{eff_channel_spec_convg_to_channel_spec} demonstrates that the proposed analog precoder $\bF^{\star}_{\text{RF}}$ and combiner $\bW^{\star}_{\text{RF}}$ are asymptotically optimal for the problem (P7), and thus are highly effective in practical mmWave MIMO-OFDM systems.

\subsection{Baseband Beamformer Design for MIMO-OFDM Systems}
In contrast to the IRS reflection matrix and analog precoder/combiner, the baseband precoder and combiner can be designed differently for each subcarrier. This flexibility greatly simplifies the generalization of the proposed baseband beamformer design from narrowband to broadband MIMO systems. That is, at each subcarrier $k \in \{0,\dots,K-1\}$, we can directly utilize the baseband beamformer design in Section \ref{JointDesign_digital} to construct the baseband combiner $\bW^\star_\text{BB}[k] \in \mathbb{C}^{N_\text{r}^\text{RF} \times N_\text{s}}$ and precoder $\bF^\star_\text{BB}[k] \in \mathbb{C}^{N_\text{t}^\text{RF} \times N_\text{s}}$ that asymptotically attain the maximum achievable spectral efficiency $R_\text{max}(\bH_\text{eff}^\star[k])$ under $\bH_\text{eff}^\star[k]=(\bW^\star_\text{RF})^\mathrm{H}\bH^\star_\text{tot}[k]\bF^\star_{\text{RF}}$. Denoting the SVD of $\bH_\text{eff}^\star[k]$ as $\bH_\text{eff}^\star[k]=\bU_\text{eff}^\star[k]\bSigma_\text{eff}^\star[k](\bV_\text{eff}^\star[k])^\mathrm{H}$, we can express $\bW^\star_\text{BB}[k]$ and $\bF^\star_\text{BB}[k]$ as
	\begin{align}
	\bW^\star_\text{BB}[k]&=[\bU^\star_\text{eff}[k]]_{:,1:N_\text{s}},\notag\\
	\bF^\star_\text{BB}[k]&=\gamma^\star[k][\bV^\star_\text{eff}[k]]_{:,1:N_\text{s}}(\bP^\star_\text{eff}[k])^{1/2},\label{FW_star_k}
	\end{align}
where $(\bP^\star_\text{eff}[k])^{1/2}=\text{diag}\left(\left[\sqrt{P^\star_1[k]},...,\sqrt{P^\star_{N_\text{s}}[k]}\medspace\right]\right)$, $P^\star_l[k]=\left(1/\eta^\star[k]-\sigma^2_{n}/|[\bSigma^\star_\text{eff}[k]]_{l,l}|^2\right)^+, \forall l \in \{1,\dots,N_\text{s}\}$, and $\eta^\star[k]$ is chosen such that $\sum_{l=1}^{N_\text{s}}P^\star_l[k]=P_\text{TX}[k]$. Similar to $\gamma^\star$ in \eqref{gamma_star}, the normalization factor $\gamma^\star[k]$ is defined as 
\begin{align}
\gamma^\star[k]=\frac{\sqrt{P_\text{TX}[k]}}{\|\bF^\star_\text{RF}[\bV^\star_\text{eff}[k]]_{:,1:N_\text{s}}(\bP^\star_\text{eff}[k])^{1/2}\|_\mathrm{F}}, \label{gamma_star_k}
\end{align}
so that the transmit power constraint is met with equality at each subcarrier, i.e., $\|\bF^\star_\text{RF}\bF_\text{BB}^\star[k]\|_\mathrm{F}^2=P_\text{TX}[k],\medspace \forall k \in \{0,\dots,K-1\}$. Since $(\bF^\star_\text{RF})^\mathrm{H}\bF^\star_\text{RF} \rightarrow \bI_{N^\text{RF}_\text{t}}$ and $(\bW^\star_\text{RF})^\mathrm{H}\bW^\star_\text{RF}\rightarrow\bI_{N^\text{RF}_\text{r}}$ as $N_\text{t},N_\text{r} \rightarrow \infty$, it follows directly from the discussions in Section \ref{JointDesign_digital} and \eqref{eff_channel_spec_convg_to_channel_spec} that the proposed hybrid beamformer $\{\bW^\star_\text{RF}\bW^\star_\text{BB}[k],\bF^\star_\text{RF}\bF_\text{BB}^\star[k]\}_{k=0}^{K-1}$ achieves spectral efficiency that approaches arbitrarily close to $\sum_{k=0}^{K-1} R_\text{max}(\bH^\star_\text{tot}[k])$ as $N_\text{t}, N_\text{r},M \rightarrow \infty$. Given that IRS-aided mmWave MIMO-OFDM systems typically employ a large number of antennas and IRS elements, it can be concluded that, even with a significantly reduced number of RF chains, the proposed hybrid beamformer can attain performance close to that of fully-digital beamformer, as will be demonstrated in Section \ref{section:simresult}.

\section{Complexity Analysis of Proposed Joint Designs} \label{section:DandCA}
In this section, we analyze the computational complexities of the proposed joint designs of IRS reflection matrix and hybrid beamformer for narrowband and broadband mmWave MIMO systems. A summary of the proposed designs are provided in Algorithm \ref{alg1} and Algorithm \ref{alg2}. Since the systems of interest typically employ a large number of antennas and IRS elements, we assume throughout the analysis that $N_\text{t}$, $N_\text{r}$, and $M$ are much greater than $N^\text{RF}_\text{t}, N^\text{RF}_\text{r}$ and $N_\text{path}^{i}, i \in \{\text{TR, TI, IR}\}$. Under this assumption, the complexities of the proposed IRS design and hybrid beamformer design for narrowband MIMO systems are given by $\mathcal{O}(N_\text{t}N_\text{r}M)$ and $\mathcal{O}(N_\text{t}N_\text{r}+N_\text{t}^\text{RF}N_\text{r}^\text{RF}N_\text{min}^\text{RF})$, respectively. Therefore, the proposed joint design for narrowband systems has the complexity of  $\mathcal{O}(N_\text{t}N_\text{r}M)$. Similarly, the complexity of the proposed joint design for MIMO-OFDM systems is calculated to be $\mathcal{O}(KN_\text{t}N_\text{r}M)$. 

Table I shows the comparison of the computational complexities of the proposed joint designs, MO-based design \cite{WANG21}, and geometric mean decomposition (GMD)-based design \cite{YING20}. By smartly leveraging the structures of frequency-flat and frequency-selective mmWave channels, the proposed designs achieve the lowest complexity for both narrowband and broadband MIMO-OFDM systems. 

\begin{algorithm}[t]
	\caption{Design of IRS Reflection Matrix and Hybrid Beamformer for Narrowband MIMO Systems}\label{alg1}
	\begin{algorithmic}[1]
		\State Compute the IRS reflection matrix $\bPhi^\star=\text{diag}(\bv^\star)$, where $\bv^\star=M\text{diag}(\ba_\text{r}(\phi^\text{r}_{\text{TI},0},\theta^\text{r}_{\text{TI},{0}})^\mathrm{H})\ba_\text{t}(\phi^\text{t}_{\text{IR},0},\theta^\text{t}_{\text{IR},0})$.
		\State Construct the analog precoder $\bF_\text{RF}^\star$ and combiner $\bW_\text{RF}^\star$ according to \eqref{analog_prec_narrow} and \eqref{analog_comb_narrow}.
		\State Obtain the baseband combiner $\bW_\text{BB}^\star=\hat{\bW}^\star_\text{BB}$ and precoder $\bF_\text{BB}^\star=\gamma^\star\hat{\bF}^\star_\text{BB}$ from \eqref{tilde_F_BB_and_W_BB^star} and \eqref{gamma_star}.
	\end{algorithmic}
\end{algorithm}

\begin{algorithm}[t]
	\caption{Design of IRS Reflection Matrix and Hybrid Beamformer for Broadband MIMO-OFDM Systems}\label{alg2}
	\begin{algorithmic}[1]
		\State Set the IRS reflection matrix as $\bPhi^\star=\text{diag}(\bv^\star)$, where $\bv^\star=M\text{diag}(\ba_\text{r}(\phi^\text{r}_{\text{TI},0},\theta^\text{r}_{\text{TI},{0}})^\mathrm{H})\ba_\text{t}(\phi^\text{t}_{\text{IR},0},\theta^\text{t}_{\text{IR},0})$.
		\State Construct the analog precoder $\bF_\text{RF}^\star$ and combiner $\bW_\text{RF}^\star$ according to \eqref{analog_prec_wideband} and \eqref{analog_comb_wideband}.
		\State Obtain the baseband combiner $\bW_\text{BB}^\star[k]$ and precoder $\bF_\text{BB}^\star[k]$ for each subcarrier $k \in \{0,\dots,K-1\}$ from \eqref{FW_star_k} and \eqref{gamma_star_k}.
	\end{algorithmic}
\end{algorithm}
\begin{table}[t!]
	\begin{center}
		\renewcommand{\arraystretch}{1.3}
		\captionsetup{labelsep=newline}
		\caption{Computational Complexities of Different Designs}
		\scalebox{0.915}{
			\begin{tabular}{| c | c | c |}
					\hline
				\text{Systems} & Design & \text{Computational Complexity}  \\
				\hline
				\multirow{3}{*}{\begin{tabular}{l}Narrowband \\
						\hspace{8pt}MIMO
				\end{tabular}} & Proposed &$\mathcal{O}(N_\text{t}N_\text{r}M)$\\ %\cline{2-3}
				& MO-based \cite{WANG21}& $\mathcal{O}(N_\text{t}N_\text{r}M+N_\text{t}N_\text{r}\text{min}(N_\text{t},N_\text{r}))$ \\ 
				&GMD-based \cite{YING20}& $\mathcal{O}(N_\text{t}N_\text{r}M+N_\text{t}N_\text{r}\text{min}(N_\text{t},N_\text{r}))$	\\ \hline
				\multirow{2}{*}{MIMO-OFDM}&  Proposed  &$\mathcal{O}(KN_\text{t}N_\text{r}M)$\\ %\cline{2-3}
				& MO-based \cite{WANG21}& $\mathcal{O}(KN_\text{t}N_\text{r}M+KN_\text{t}N_\text{r}\text{min}(N_\text{t},N_\text{r}))$ \\
				& GMD-based \cite{YING20}& $\mathcal{O}(KN_\text{t}N_\text{r}M+KN_\text{t}N_\text{r}\text{min}(N_\text{t},N_\text{r}))$ \\ 
				\hline
		\end{tabular}}
	\end{center}
	{\label{table1}}
\end{table} 

	\section{Simulation Results} \label{section:simresult}
	In this section, we present simulation results to demonstrate the effectiveness of the proposed joint designs of IRS reflection matrix and hybrid beamformer for narrowband and broadband MIMO systems. We assume that the TX equipped with a UPA of $N_\text{t}=8 \times 8=64$ antennas communicates to the RX with a UPA of $N_\text{r}=4 \times 4=16$ antennas. The number of RF chains at the TX and RX is set to be $N_\text{t}^\text{RF}=N_\text{r}^\text{RF}=4$. Unless otherwise specified, we assume that the IRS is equipped with a UPA of $M=16 \times 16=256$ passive elements and that $\bH_i$ contains $N_\text{path}^{i}=8$ propagation paths, $\forall i \in\{\text{TR, TI, IR}\}$. The distance $d_\text{TR} \in [ d_\text{TI}+d_\text{IR}-10\medspace\text{m}, d_\text{TI}+d_\text{IR}\medspace\text{m})$ between the TX and RX follows a uniform distribution over its range given $d_\text{TI}$ and $d_\text{IR}$, each of which are uniformly distributed over $[50\medspace\text{m}, 60\medspace\text{m}]$ and $[10\medspace\text{m}, 20\medspace\text{m}]$. The distance-dependent path loss $PL(d_i)$ is modeled as 
\begin{align}
PL(d_i)\medspace[\text{dB}]=\alpha+10\beta\log_{10}(d_i)+\xi, \label{PL_model}
\end{align} 
where $\xi\sim\cN(0,\sigma^2)$. According to the experimental data for 28 GHz channels in \cite{AKDENIZ14}, the parameters in \eqref{PL_model} are set to be $\alpha=61.4, \beta=2, \sigma=5.8\medspace\text{dB}$ for a line-of-sight (LOS) path of $\bH_i$, and $\alpha=72.0, \beta=2.92, \sigma=8.7\medspace\text{dB}$ for its non-line-of-sight (NLOS) paths. To evaluate the effectiveness of different IRS reflection matrix designs more accurately, we assume that each path of $\bH_\text{TR}$ is a NLOS path that passes through tinted-glass walls to experience an additional penetration loss of $40.1\medspace\text{dB}$ \cite{ZHAO13}. The element spacing, noise power, and number of data streams are each set to be $d=\lambda/2, \sigma^2_{n}=-91 \medspace\text{dBm}$, and $N_\text{s}=4$. Lastly, all the simulation results are averaged over 10,000 channel realizations.

	\subsection{Narrowband MIMO Systems}\label{subsection:sim_result_narrowband}
\begin{figure}[!t]
	\centering
	\includegraphics[width=7.5cm]{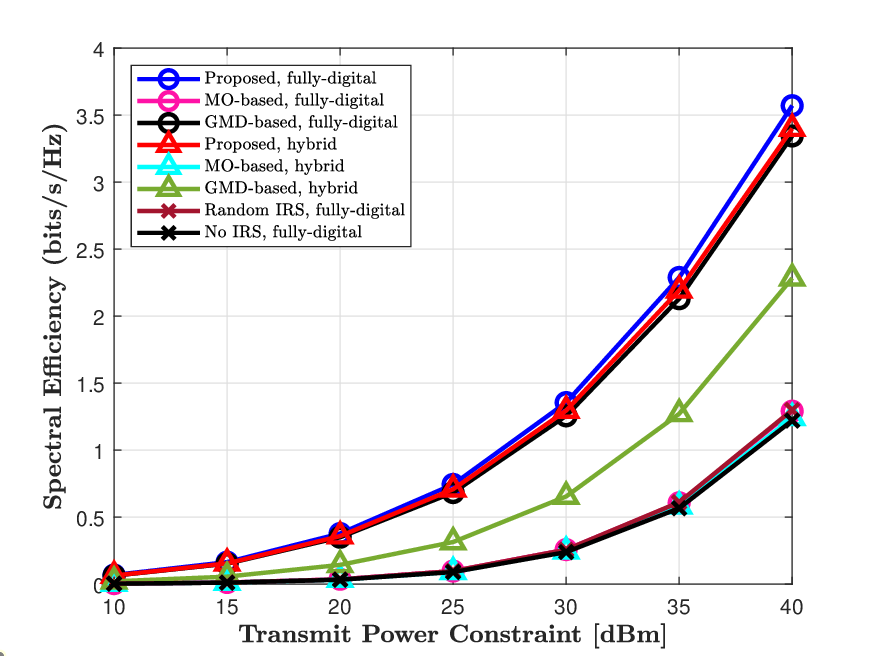}
	%   % where an .eps filename suffix will be assumed under latex,
	%   % and a .pdf suffix will be assumed for pdflatex
	\caption{Spectral efficiency achieved by different designs as a function of the transmit power constraint $P_\text{TX}$.}\label{Fig2}
\end{figure}

In this subsection, we evaluate the performance of the proposed joint design of IRS reflection matrix and hybrid beamformer for narrowband MIMO systems. Fig. \ref{Fig2} shows the spectral efficiency achieved by the proposed design as a function of the transmit power constraint $P_\text{TX}$. We also plotted in Fig. \ref{Fig2} the performances of when the phase shift of each IRS element is randomly selected from $[0\medspace \text{rad},2\pi \medspace\text{rad})$ and when there is no IRS, each of which are labeled as ``Random IRS'' and ``No IRS''. The results show that, with hybrid beamforming, the proposed design outperforms MO-based and GMD-based designs that require higher complexities. In addition, the spectral efficiency of the proposed design with hybrid beamforming is almost the same as that with fully-digital beamforming, demonstrating the effectiveness of the hybrid beamformer design presented in Section \ref{sec:jointdesign}. In contrast, there is a significant performance gap between GMD-based design with hybrid beamforming and that with fully-digital beamforming. This is primarily because the design aims to construct hybrid beamformers that achieve low BERs instead of high achievable rates. It can also be observed that there are negligible differences among the spectral efficiency achieved by MO-based design, Random IRS, and No IRS. This indicates that, in order to obtain performance gains from IRS, it is important to take into account the presence of the direct channel and design the IRS reflection matrix carefully.

	\begin{figure}[!t]
	\centering
	\includegraphics[width=7.5cm]{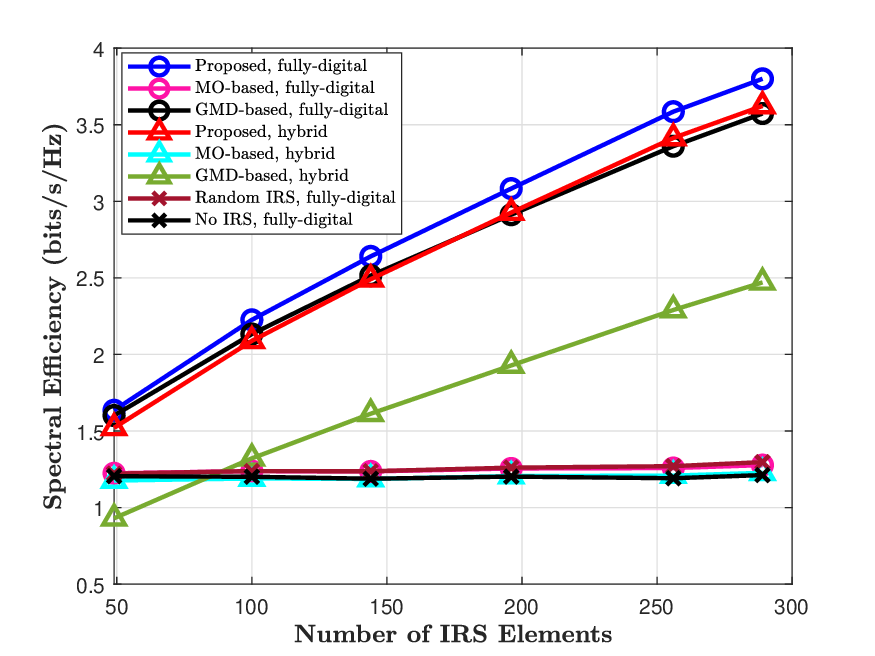}
	%   % where an .eps filename suffix will be assumed under latex,
	%   % and a .pdf suffix will be assumed for pdflatex
	\caption{Spectral efficiency achieved by different designs as a function of the number of IRS elements $M$.}\label{Fig3}
\end{figure}

To investigate the impact that the number of IRS elements $M$ has on the proposed design and other benchmarks, the spectral efficiency of different designs is plotted as a function of $M$ in Fig. \ref{Fig3}, where $P_\text{TX}=40\medspace\text{dBm}$ and $M^h=M^v=\sqrt{M}$. The figure shows that, as discussed in Section \ref{sec:IRS_design_singletap}, the spectral efficiency achieved by the proposed design significantly increases with $M$. In contrast, the performances of Random IRS and MO-based design do not improve monotonically with $M$. This indicates that, regardless of how many IRS elements are utilized, the judicious design of IRS reflection matrix is necessary to achieve spectral efficiency gains from IRS. Furthermore, the proposed design with hybrid beamforming performs just as well as that with fully-digital beamforming at all values of $M$, while GMD-based design requires fully-digital beamformers to perform comparably to the proposed design. This result demonstrates that the proposed joint design is highly suitable for mmWave MIMO systems, where fully-digital beamforming necessitates a large number of RF chains and thus is prohibitively expensive.

\begin{figure}[!t]
	\centering
	\includegraphics[width=7.5cm]{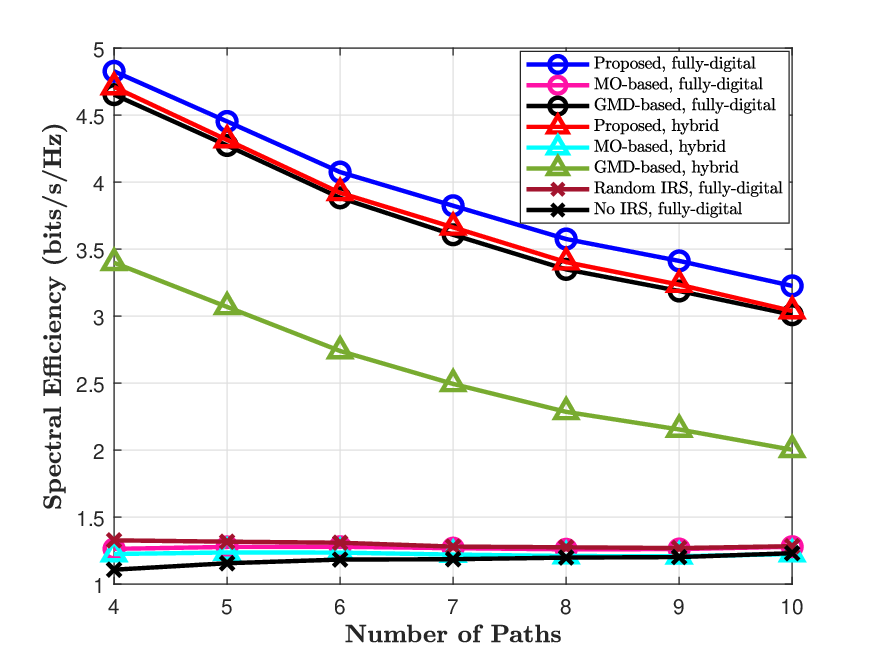}
	%   % where an .eps filename suffix will be assumed under latex,
	%   % and a .pdf suffix will be assumed for pdflatex
	\caption{Spectral efficiency achieved by different designs as a function of the number of paths $N_\text{path}$.}\label{Fig4}
\end{figure}
Fig. \ref{Fig4} plots the spectral efficiency of different designs against the number of paths $N_\text{path}=N_\text{path}^{i}, i \in \{\text{TR, TI, IR}\}$, when  $P_\text{TX}=40\medspace\text{dBm}$. As shown in the figure, the spectral efficiency of the proposed design and GMD-based design increases as $N_\text{path}$ decreases, while that of Random IRS and MO-based design does not change significantly according to $N_\text{path}$. This implies that the proposed design is well suited to mmWave systems in which communications are generally performed in environments with low scattering and small number of paths.

\begin{figure}[!t]
	\centering
	\includegraphics[width=7.5cm]{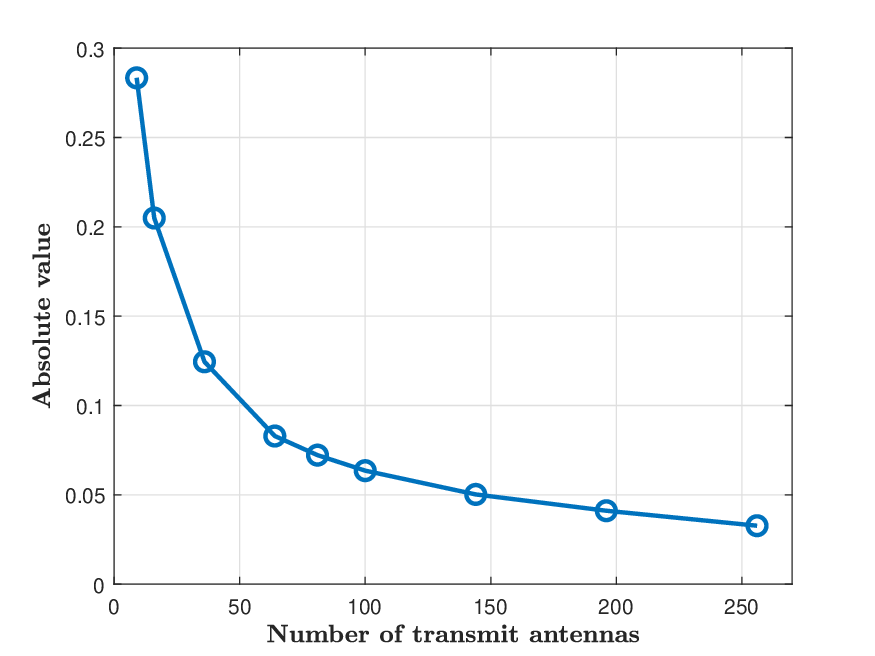}
	%   % where an .eps filename suffix will be assumed under latex,
	%   % and a .pdf suffix will be assumed for pdflatex
	\caption{Absolute value of an element of $(\bA_\text{t}^\text{TR})^\mathrm{H}\bA_\text{t}^\text{TI}$ as a function of the number of transmit antennas $N_\text{t}$.}\label{Fig_asymptot}
\end{figure}

\begin{figure}[!t]
	\centering
	\includegraphics[width=7.5cm]{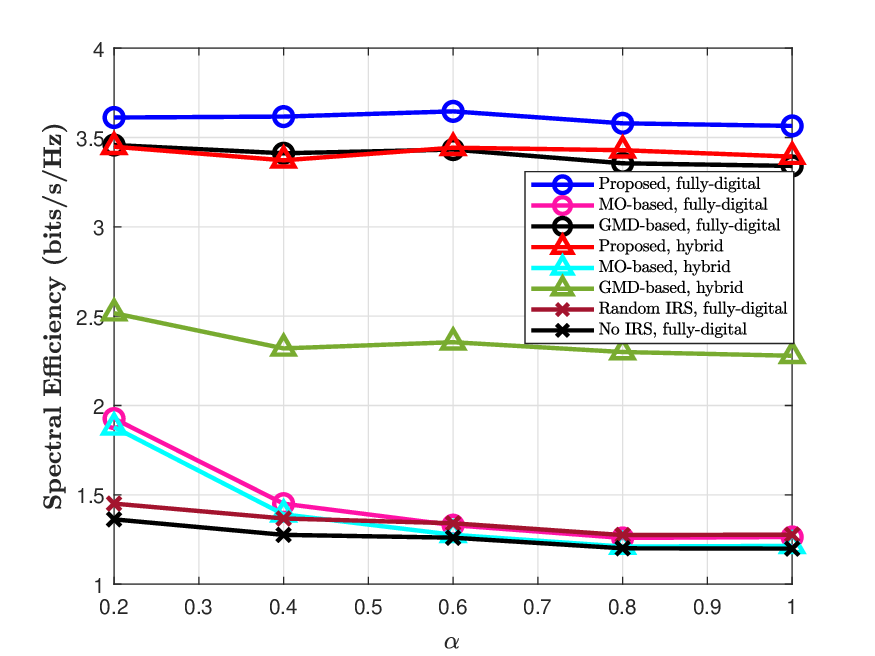}
	%   % where an .eps filename suffix will be assumed under latex,
	%   % and a .pdf suffix will be assumed for pdflatex
	\caption{Spectral efficiency achieved by different designs as a function of the angular range parameter $\nu$.}\label{Fig_ang_spread}
\end{figure}

Since the effectiveness of the proposed joint design depends on how uncorrelated the array response vectors corresponding to different paths are to each other, we evaluate in Fig. \ref{Fig_asymptot} the absolute value of an element of the matrix $(\bA_\text{t}^\text{TR})^\mathrm{H}\bA_\text{t}^\text{TI}$ as a function of $N_\text{t} $. The figure shows that each element of the matrix rapidly approaches $0$ as $N_\text{t}$ increases, demonstrating that the result in \eqref{array_matrix_convergence} is highly relevant to mmWave systems where large antenna arrays are typically deployed. In addition, Fig. \ref{Fig_ang_spread} shows the spectral efficiency of different designs when the azimuth and elevation AoAs/AoDs are each uniformly distributed over $[0 \medspace\text{rad},2\nu\pi\medspace\text{rad})$ and $[0\medspace\text{rad},\nu\pi\medspace\text{rad})$, where $\nu \in [0,1]$ and the transmit power constraint is set as $P_\text{TX}=40\medspace\text{dBm}$. The result shows that the spectral efficiency achieved by the proposed joint design does not change significantly according to $\nu$. Such consistency can be attributed to the fact that the array response vectors corresponding to different paths are weakly correlated in the systems of interest, as implied by Fig. \ref{Fig_asymptot}. Also, when hybrid beamforming is used, the proposed design outperforms all the other benchmarks at each value of $\nu$. This demonstrates that the design can be effectively employed in practical IRS-aided mmWave systems where a large number of antennas makes hybrid beamforming an economical and efficient alternative to fully-digital beamforming.

\begin{figure}[!t]
	\centering
	\includegraphics[width=7.5cm]{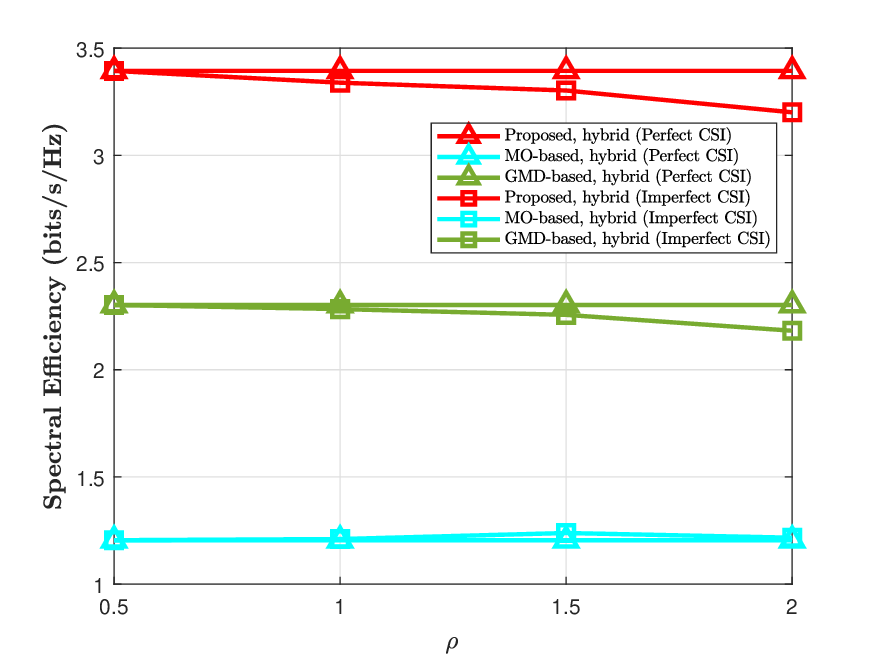}
	%   % where an .eps filename suffix will be assumed under latex,
	%   % and a .pdf suffix will be assumed for pdflatex
	\caption{Spectral efficiency achieved by different designs as a function of the estimation error parameter $\rho$.}\label{Fig_channel_est}
\end{figure}

In order to examine how the channel estimation error affects the performance of the proposed design, we plot in Fig. \ref{Fig_channel_est} the spectral efficiency achieved by different designs when the estimated channel $\hat{\bH}_i$ is used to construct the IRS reflection matrix and hybrid beamformer. For each $i \in \{\text{TR, TI, IR}\}$, the estimated channel $\hat{\bH}_i$ is given by 
	\begin{align}
	\hat{\bH}_i=\sum_{q=0}^{N_\text{path}^i-1}\hat{\alpha}_{i,q}\ba_\text{r}(\hat{\phi}^\text{r}_{i,q},\hat{\theta}^\text{r}_{i,q})\ba_\text{t}(\hat{\phi}^\text{t}_{i,q},\hat{\theta}^\text{t}_{i,q})^{\mathrm{H}},
	\end{align}
	where the estimated complex gain and azimuth AoA of the $q$-th path are respectively denoted by $\hat{\alpha}_{i,q}=(1+\delta)\alpha_{i,q}$ and $\hat{\phi}^\text{r}_{i,q}=\phi^\text{r}_{i,q}+\delta \text{ [deg]}$. The estimation noise $\delta$ is assumed to be uniformly distributed over $[-\rho,\rho]$ \cite{WANG_TWC}, and other estimated AoAs/AoDs are defined in a similar manner. The transmit power constraint is fixed at $P_\text{TX}=40\text{ dBm}$. The result in Fig. \ref{Fig_channel_est} shows that the proposed design with imperfect CSI outperforms the benchmarks with perfect CSI at each value of $\rho$. In addition, the presence of the channel estimation error does not lead to significant degradation in the performance of the proposed design. It can thus be concluded that the proposed design is highly robust against the channel estimation error that inevitably occurs in practical communication systems.

	\begin{figure}[!t]
	\centering
	\subfloat[$K=16$]{
		\includegraphics[width=7.5cm]{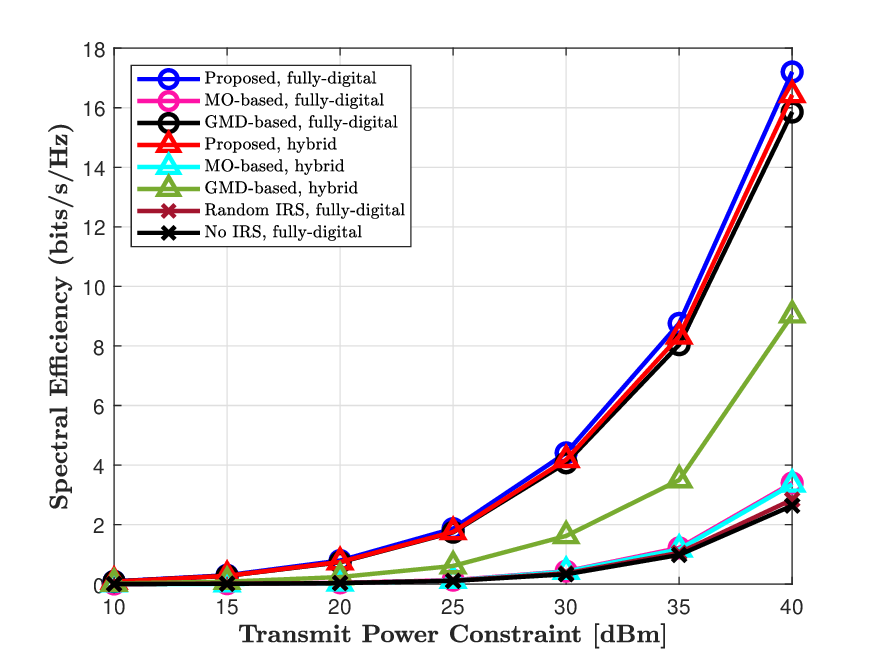}
		\label{TXP_16_subc}
	}\\
	\subfloat[$K=64$]{
		\includegraphics[width=7.5cm]{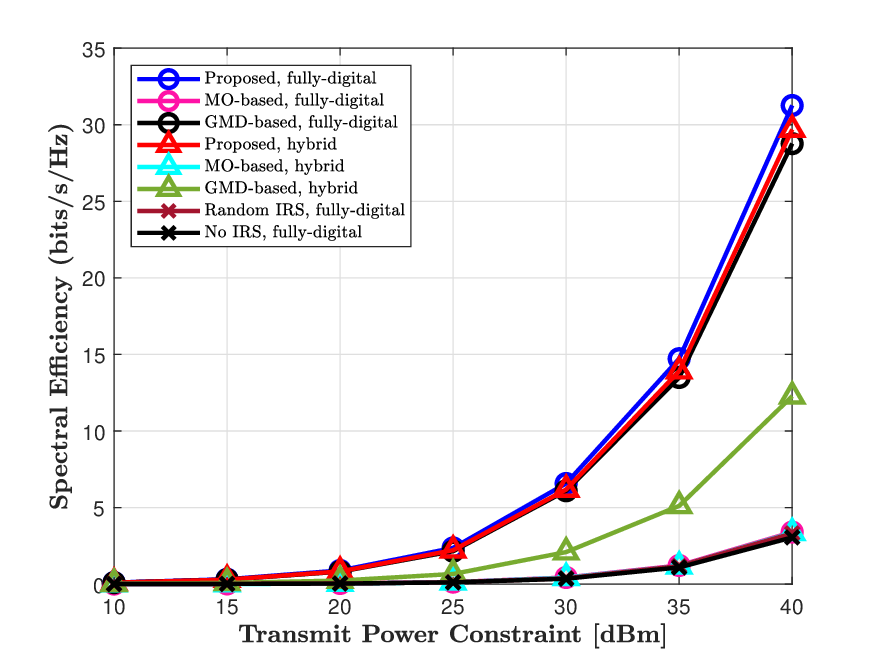}
		\label{TXP_64_subc}
	}
	\caption{Spectral efficiency of different designs in MIMO-OFDM systems with $K$ subcarriers as a function of the transmit power constraint $P_\text{TX}$.}
	\label{Fig6}
\end{figure}

	\subsection{Broadband MIMO-OFDM Systems} \label{subsection:sim_result_broadband}
In this subsection, we investigate the performance of the proposed joint design of IRS reflection matrix and hybrid beamformer for broadband MIMO-OFDM systems. We assume that equal power is allocated to each subcarrier, i.e., $P_\text{TX}[k]=P_\text{TX}/K,\medspace \forall k \in \{0,\dots,K-1\}$. In Figs. \ref{Fig6} (a) and (b), the spectral efficiency that different designs achieve in MIMO-OFDM systems with $K=16$ and $K=64$ subcarriers is plotted as a function of $P_\text{TX}$. The figures show that the proposed design outperforms all the other benchmarks at both values of $K$, regardless of whether fully-digital or hybrid beamforming architectures are utilized. Also, the proposed design with hybrid beamforming achieves spectral efficiency very close to that of the design with fully-digital beamforming. In contrast, a significant performance gap exists between GMD-based design with fully-digital and hybrid beamforming, indicating that the proposed hybrid beamformer design is superior in terms of maximizing spectral efficiency. As demonstrated by the results in Fig. \ref{Fig6}, the proposed joint design, which carefully exploits the angular sparsity of frequency-selective mmWave channels, is capable of providing substantial improvements in the spectral and energy efficiency of MIMO-OFDM systems.

	\section{Conclusions} \label{sec:conclusions}
In this paper, we studied narrowband and broadband IRS-aided mmWave MIMO systems with hybrid beamforming architectures. We first formulated the problem of designing the IRS reflection matrix and analog beamformer for narrowband MIMO systems into the effective channel design problem. By leveraging the sparse-scattering structure and large dimension of mmWave channels, we developed the joint design of IRS reflection matrix and hybrid beamformer for narrowband MIMO systems. We generalized the proposed joint design for narrowband MIMO systems to broadband MIMO-OFDM systems by carefully exploiting the sparsity of frequency-selective mmWave channels in the angular domain. Simulation results demonstrated that the proposed designs can provide the systems of interest with significant spectral efficiency gains and outperform the existing state-of-the-art designs while requiring lower computational complexity. Interesting future research directions include the investigation of the IRS reflection matrix and beamformer design that minimizes the inter-user interference in IRS-aided multi-user systems with hybrid beamforming architectures.

	% or
	

	\appendices
	\section{Proof of Lemma 1} \label{appendix_A}
	Let $j \in \{0,\dots,N_\text{path}^\text{TI}-1\}$. Define a unitary matrix $\bU$ and a diagonal matrix $\bLambda$ such that $\bH_\text{IR}^\mathrm{H}\bH_\text{IR}=\bU\bLambda\bU^\mathrm{H}$, where $[\bLambda]_{m,m}$ is necessarily an eigenvalue of $\bH_\text{IR}^\mathrm{H}\bH_\text{IR}$, $\forall m \in \{1,\dots,M\}$. Since it holds for each $m$ that $[\bLambda]_{m,m} \leq \lambda_0(\bH_\text{IR}^\mathrm{H}\bH_\text{IR})$, $\norm{\bq_j}_2^2$ can be bounded as 
	\begin{align}
\norm{\bq_j}^2_2&=\ba_\text{r}(\phi^\text{r}_{\text{TI},j},\theta^\text{r}_{\text{TI},j})^\mathrm{H}\bPhi^\mathrm{H}\bH_\text{IR}^\mathrm{H}\bH_\text{IR}\bPhi\ba_\text{r}(\phi^\text{r}_{\text{TI},j},\theta^\text{r}_{\text{TI},j})=\sum_{m=1}^{M}[\bLambda]_{m,m}|[\bU^\mathrm{H}\bPhi\ba_\text{r}(\phi^\text{r}_{\text{TI},j},\theta^\text{r}_{\text{TI},j})]_m|^2 \notag\\
	&\leq \lambda_0(\bH_\text{IR}^\mathrm{H}\bH_\text{IR})\norm{\bU^\mathrm{H}\bPhi\ba_\text{r}(\phi^\text{r}_{\text{TI},j},\theta^\text{r}_{\text{TI},j})}_2^2=\lambda_0(\bH_\text{IR}^\mathrm{H}\bH_\text{IR}), \label{bound_q_j}
	\end{align}
where the last equality follows from $\norm{\bPhi\ba_\text{r}(\phi^\text{r}_{\text{TI},j},\theta^\text{r}_{\text{TI},j})}_2^2=1$. We now prove the equality condition of the inequality (\ref{inequality_lemma1}) in Lemma 1. If $\bH_\text{IR}^\mathrm{H}\bH_\text{IR}\bPhi\ba_\text{r}(\phi^\text{r}_{\text{TI},j},\theta^\text{r}_{\text{TI},j})=\lambda_0(\bH_\text{IR}^\mathrm{H}\bH_\text{IR})\bPhi\ba_\text{r}(\phi^\text{r}_{\text{TI},j},\theta^\text{r}_{\text{TI},j})$, $\norm{\bq_j}^2_2$ can be written as
	\begin{align}
	\norm{\bq_j}^2_2&=\lambda_0(\bH_\text{IR}^\mathrm{H}\bH_\text{IR})\norm{\bPhi\ba_\text{r}(\phi^\text{r}_{\text{TI},j},\theta^\text{r}_{\text{TI},j})}_2^2=\lambda_0(\bH_\text{IR}^\mathrm{H}\bH_\text{IR}).
	\end{align}
	This completes the proof of Lemma 1. \QEDB

	\ifCLASSOPTIONcaptionsoff
	\newpage
	\fi

	
	
	%\bibliographystyle{IEEEtran}
	\bibliographystyle{IEEEtran}
	\bibliography{HBFwithIRS_reference_2col}


	
	%\vfill
	

\end{document}